\documentclass[11pt,twocolumn]{article}

\usepackage[a4paper,top=3.5cm,width=15.5cm,vscale=.75,head=20.3pt]{geometry}	

\usepackage{graphicx}
\usepackage{dcolumn}
\usepackage{bm}
\usepackage{amsmath,amsfonts}

\usepackage{url}
\usepackage{hyperref} 
\usepackage{multirow}

\usepackage[natbibapa]{apacite}
\let\cite\citep
\begin{document}

\title{Modeling synchronization in human musical rhythms using Impulse Pattern Formulation (IPF)}

\author{Simon Linke\textsuperscript{1),2)}, Rolf Bader\textsuperscript{1)}, Robert Mores\textsuperscript{2)}\\
\textsuperscript{1)}Institute of Systematic Musicology,  University of Hamburg, 20354 Hamburg, Germany\\
\textsuperscript{2)}Hamburg University of Applied Sciences, 22081 Hamburg, Germany\\
 linke@mt.haw-hamburg.de}

\twocolumn[
  \begin{@twocolumnfalse}
    \maketitle
    
    \begin{abstract}
      When musicians perform in an ensemble, synchronizing to a mutual pace is the foundation of their musical interaction. Clock generators, e.g., metronomes, or drum machines, might assist such synchronization, but these means, in general, will also distort this natural, self-organized, inter-human synchronization process. In this work, the synchronization of musicians to an external rhythm is modeled using the Impulse Pattern Formulation (IPF), an analytical modeling approach for synergetic systems motivated by research on musical instruments. Nonlinear coupling of system components is described as the interaction of individually propagating and exponentially damped impulse trains. The derived model is systematically examined by analyzing its behavior when coupled to numerical designed and carefully controlled rhythmical beat sequences. The results are evaluated by comparison in the light of other publications on tapping. Finally, the IPF model can be applied to analyze the personal rhythmical signature of specific musicians or to replace drum machines and click tracks with more musical and creative solutions.
    
    \end{abstract}

  \end{@twocolumnfalse}
  ]

\section{\label{sec:intro} Introduction}

Drum machines or computer-based click tracks are standard in today's music and used in recording studios and live concerts. These solutions are convenient for musicians and recording engineers, but they are usually perceived as a bit artificial (or even dull) because of the fixed beat and the lack of minor imperfections that inevitably occur when human musicians perform independently. Those deviations may be intended due to musical expression (e.g., agogic accents) or unintended due to human imperfections. Nevertheless, there are some musical genres or certain musical pieces where a perfect beat with no deviations seems to be the most appropriate (see, e.g., \citet{Datseris.2019}).

A common approach to add a human feel to electric or digital generated rhythms is to add slight random deviations to every beat, but according to \citet{Hennig.2012} the results are rarely satisfying. Many authors state that the human fluctuations can be explained as pink 1/f-noise (see, e.g., \citet{Coey.2016,Gilden.2001,vanOrden.2003}). More general \citet{Hennig.2011} investigate long-range correlated noise based on the power law $1/f^B $, where $0.2<B<1.3$ leads to satisfactory results.  

However, not every author agrees that correlated noise is the best solution: \textit{``Variation in rhythmical movement can not be modeled as an independent random process (noise), and is consider to be part of the intrinsic dynamics of a movement system, which is consider to be a dynamical system''}\citep[p. 383]{Yamada.1995}.

Such a dynamical system, which also relies on long-range correlation, has been proposed by \citet{Haken.1985}. It is a nonlinear model based on harmonic oscillators mutually coupled to a periodical potential, responsible for phase changes. Due to its nonlinear approach, this model reproduces sudden phase changes and hysteresis, which occur when two systems synchronize to a common tempo. Further, this model can be extended to explain polyrhythmic behavior \cite{Haken.1996}.

In live music performances, often two or more musicians play music together. They aim to mutually synchronize their pace to each other. All musicians may slightly variate their tempo, but they still have to agree to a common pace. This process cannot be satisfyingly investigated by the approaches mentioned above (e.g., \cite{Hennig.2011}), where musicians still have to synchronize to a fixed metronome-like pattern. Whether this pattern is modulated with long-range corrected fluctuation or not, it does not adapt to a musician's tempo. However, this mutual coupling can be simulated by the model of \citet{Haken.1985}, and further, other nonlinear models could be transferred. For instance, feedback-loops with added noise have been studied in terms of mode-hopping \cite{DHuys.2016} or coherence resonance \cite{Just.2016}.

In this work, a metronome-like algorithm is developed. It reacts to changes in the tempo of musicians, similar to what a real musician would do. It can reproduce the tempo of a musician but also show minor imperfections, responsible for a more human-like feel. By varying a limited set of system parameters, it can be used to replace a real musician or to act as a creative character on its own.

The core of this modeling approach is the Impulse Pattern Formulation (IPF), described in detail in Section \ref{sec:IPF}. It is a nonlinear recursive equation derived by \citet{Bader.2013} primarily to model musical instruments. The IPF is based on the idea of coupling different parts of musical instruments, but it can also describe all kinds of self-organized synchronization processes, like in this work, musicians synchronizing to a common tempo. Due to its recursive nature, the IPF reproduces small fluctuations in time and transients. Its general nonlinear approach allows sudden phase transitions to be modeled, similar to \citet{Haken.1985}. Like many other nonlinear recursive equations (e.g., the logistic map or the sine circle map), the IPF is a relatively simple equation capable of causing complex, chaotic behavior. Thus it usually causes low computational costs, which allows straightforward real-time coupling to human musicians.

\section{Impulse Pattern Formulation}\label{sec:IPF} 

The Impulse Pattern Formulation is an abstract top-down modeling approach to describe the transient behavior of arbitrary, coupled systems. It was introduced initially by \citet{Bader.2013} to model any musical instrument with one universal set of mathematical equations, to allow a comparison of different musical instruments, focusing on stable tone production and transient behavior.

\begin{figure*}[htb]
  \begin{center}
  \includegraphics[width=.75 \linewidth]{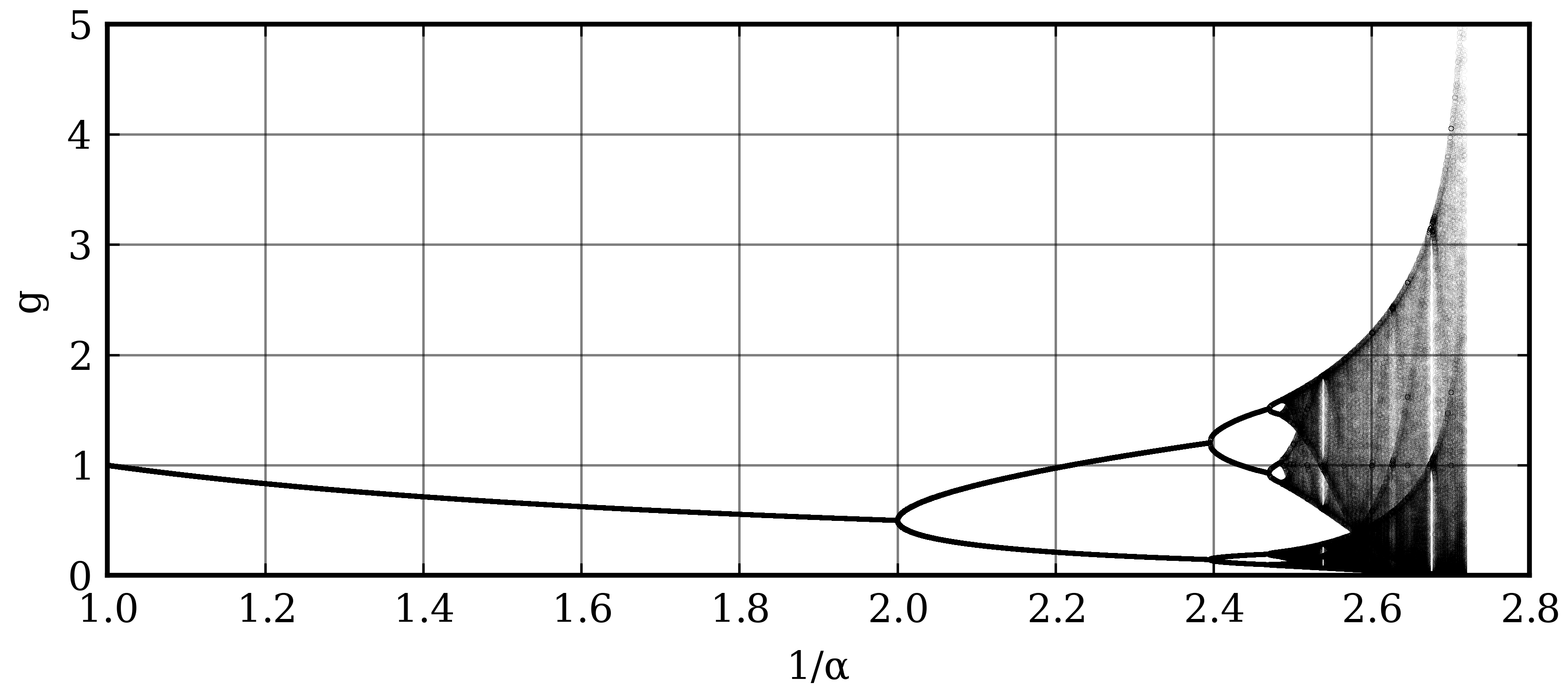} 
  \caption{Bifurcation  scenario  of the IPF with one  reflection  point $\alpha$}    
  \label{fig:Bif}
  \end{center}
\end{figure*}

A musical instrument is often described as a generator acting on a resonator (e.g., \citet{Fletcher.2010}). The IPF describes them more general as self-organized systems: A musical instrument is a system acting upon itself, consisting of mutually coupled subsystems, possibly even interacting backward. The subsystems are excited by certain impulses, which is evident for plucked-string or percussion instruments, but it is also true for sustained instruments like, e.g., violins or brass instruments (see, e.g., \cite{Linke.2019c,Bader.2013}).

To make this a bit more ostensive, we choose the simple example of a reed instrument, e.g., a saxophone. It consists of two subsystems: a reed and a tube. The reed sends out impulses that are answered by the tube. According to \citet[p.286]{Bader.2013}, the answer is just a callback of the impulses sent out by the reed. As the back-traveling impulses have an impact on the reed, the change of the system is caused by the system itself:
\begin{equation}
  \frac{\partial \bar{g}}{\partial t}=\frac 1 \alpha \bar{g} ,
\end{equation}
where $\alpha$ is the strength of the back-traveling impulse. In the chosen example of a reed instrument, this is related to the playing pressure. The system state is represented by $\bar{g}$ and is related to the amplitude and the periodicity of a signal (further explanations are given by \citet{Linke.2019c}).

Impulses need a certain amount of time to travel from one subsystem to another, during which they get exponentially damped. Thus, \citet[pp. 286-291]{Bader.2013} deduces the IPF as a recursive Equation:
\begin{equation}
  g_+=g-ln \left( \frac 1 \alpha \left( g- \sum^n_{k=1} \beta_k e^{g-g_{k-}} \right) \right) , \label{eq:IPF}
\end{equation}

where the number $k$ refers to the number of subsystems and $\beta_k$ to the related specific strengths. $g_+$ is the subsequent system state, and $g_{k-}$ are the precedent states. There is no precise time interval between those system states, as it is the time until a new event occurs. Thus, when modeling musical instruments, this is usually the period of the fundamental frequency $T_0$. 

Choosing an initial value $g_0$, Equation \eqref{eq:IPF} can be iteratively calculated. Depending on the system parameters $\alpha$ and $\beta_k$, the IPF can diverge or converge to a stable limit after several iteration steps. Further, the IPF can show chaotic behavior like bifurcations. The bifurcation scenario can be quite complex. The simplest case ($\beta_k=0 \; \forall k \in \mathbb{N}$) is shown in Fig. \ref{fig:Bif}. Here, 2500 iteration steps were performed, and the last 500 values are plotted in dependency of $\alpha$. 

The figure becomes more vivid when referring again to the given example of a reed instrument: Low playing pressure, resulting in high values $1/\alpha$, is represented on the right side of the chart. The shown unstable behavior results in noisy sounds. Increasing the playing pressure $\alpha$ results in bifurcations. Here multiple frequencies can be heard at the same time. Further increase of pressure leads to stable states resulting in regular periodic motion. The point of this sudden transition is called the first bifurcation point $\alpha_c$ An in-depth mathematical discussion of the boundaries and limits of the IPF has been given by \citet{Linke.2019b}.

The IPF can deal with the complex geometry of an instrument with little mathematical effort, but due to its general approach, it can easily be transferred to any other coupled system. The model can be further detailed by using more reflection points $\beta_k$. Such scalability is possible within musical instruments but also between entities of instruments, e.g., \citet{Linke.2021} applied the IPF to the coupling of a zither and a supporting table. In Section \ref{sec:Application} this transfer is taken even further. The IPF is used to model a musician synchronizing with the tempo of another musician or a metronome. Due to its recursive formulation, the IPF is an appropriate method to explain this problem, as it inevitably reproduces time series, including the transient behavior. 

\section{Applying the IPF on beat synchronisation}\label{sec:Application} 

\subsection{Deriving an IPF model}\label{sec:IPFmodel} 
Modeling the rhythmical synchronization between two musicians means modeling their neuronal processes.  In the past, synchronization and spatio-temporal bifurcation scenarios in neuronal networks have been successfully simulated and discussed using nonlinear recursive equations (see, e.g., \cite{Kozma.2015}). Thus, it is reasonable that the IPF can be applied, too. However, modeling the state of every active neuron would need a vast number of $\beta_k$ and thus a lot of computational resources. \citet{Haken.1996} states that when explaining human movement, usually an immense number of microscopic components (e.g., muscles or neurons) relates to macroscopic behavior. Even though this relation is often not understood, synergetic behavior can be modeled with a minimal set of order parameters. According to \citet{Bader.2013}, this is also true when applying the IPF: First, the point of observation can be freely chosen. Here, the reaction of a musician to a second musician should be modeled, so the first musician is chosen to represent the point of observation. Then, the second musician is just represented by the time-dependent input parameter $\alpha$. Secondly, the number of observed subsystems can be reduced if a less detailed observation of the overall system is sufficient to answer a specific research question. In this study, the overall behavior of the musician should be modeled rather than synchronization processes in the brain. Thus, it might be adequate to represent the musician just by a single subsystem for a first overview. Then the most simple form of the IPF can be used:

\begin{equation}
  g_+=g-ln \left( \frac g \alpha  \right) , \label{eq:IPF_simple}
\end{equation}

where $\alpha$ corresponds to the tempo of the musician and $g$ is the related tempo of the IPF. According to \cite{Linke.2019b}, Equation \eqref{eq:IPF_simple} converges against $g=\alpha$, when $\alpha$ is chosen in range sufficient for stable behavior. Thus, the tempo of the IPF converges against the tempo of the musician.

The IPF is an iterative equation, and according to Section \ref{eq:IPF}, the time between two iteration steps is not fixed. When modeling musical instruments, it refers to the fundamental frequency. In the given example of rhythmical synchronization, it refers to the period between two beats. Every time eighths notes are played, a new value $g$ is calculated whereby the new $g$ determines the length of the eighth note, and thus the period until the IPF is calculated again.

To make use of Equation \eqref{eq:IPF_simple}, the parameter $\alpha$ has to be calculated based on the tempo of the coupled musician. Therefore the time between two succeeding beats of the musician $T_{M,i}$ is measured. As the IPF relies on eighth notes, but it is unclear which note value relates to $T_{M,i}$, the error $\Delta T$ between the tempo of the IPF and the musician's tempo can be calculated using the modulo operation:

\begin{equation}
  \Delta T \equiv T_{M,i}\ (\textrm{mod}\ g) \label{eq:modulus}
\end{equation}

Thus, $\alpha$ is determined by the system state $g$ corrected by its error:
\begin{equation}
  \alpha=g+\Delta T
\end{equation}

A further restriction has to be made, as when the musician plays dotted notes or triplets, the IPF should synchronize to the fundamental meter. A straightforward solution to this problem is to ignore $\Delta T$, which are longer than a thirty-second note. This approach is also reasonable from a physical point of view, as, e.g., coupled oscillators do only synchronize if their eigenfrequencies are sufficiently similar (see, e.g., \cite{Pikovsky.2001}). Another alternative approach to this problem is discussed in Section \ref{sec:center120bpm}.

Finally, the parameters $g$ and $\alpha$ must be scaled according to the stable region of the IPF described in detail by \citet{Linke.2019b}. Fast tempos relate to small values of $\alpha$ resp. $g$. Then, a maximum tempo relates to the first bifurcation point $\alpha_c$ of the IPF. The IPF is not capable of reproducing faster tempos as it becomes chaotic and finally diverges. This maximum tempo is not restricted by the motoric and technical skills of a musician. It must be produced by the IPF only. A coupled musician can play half-time, quarter-time, or even lower due to the above mentioned modulo operator. Thus, acoustical thoughts can motivate the maximum tempo: Frequencies above $20~Hz$ are perceived as fused sounds, while lower frequencies are perceived as single beats. Thus, playing with $1200~bpm$ (beats per minute) would result in a low, buzzing sound instead of a distinguishable rhythm. Consequently, playing a quarter speed of $300~bpm$ would result in a buzzing sound when playing sixteenth notes, which seems to be the maximum tempo, that can be applied in a meaningful musical manner. If $\tau$ is a tempo in $bpm$ based on quarter notes, 

\begin{equation}
  \alpha,g=\frac{\frac{300~bpm}{\tau}-1}{18}+0.5 \ , \label{eq:scaling_simple}
\end{equation}

would related $300~bpm$ to $\alpha_c$. Further, if the tempo is $30~Hz$, $\alpha=1$ resp. $g=1$. Hence, the IPF is restricted to the unit interval for most tempi.

\subsection{Investigating the system behavior}\label{sec:IPFBehavior} 

\begin{figure*}[htb]
  \begin{center}
  \includegraphics[width=.75 \linewidth]{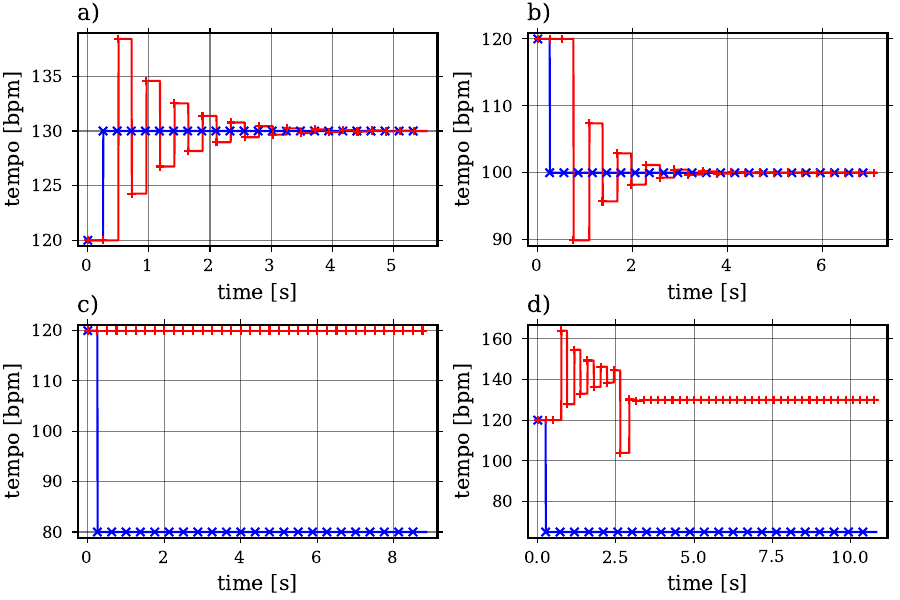} 
  \caption{IPF reacts to four different step changes in tempo. Blue lines refer to the tempo of the click track. Red lines correspond to the tempo of the IPF. The single crosses represent single beats. Audio examples of these four cases can be found at \cite{Linke.2021b}}
  \label{fig:Example}

  \end{center}
\end{figure*}

In the past, a variety of studies were conducted where participants tap along audible presented rhythmical sequences (e.g.,\cite{michon.1967,Repp.2001,Coey.2016,Handel.1981}). Applying similar sequences to the IPF allows a comparison with the literature and thus a first comparison between the IPF and human behavior. Of course, this approach does not reflect an entire synchronization process (like, e.g., described by \citet{Pikovsky.2001}), where the IPF and a musician adapt themselves to a collective tempo. Here the synchronization to a linear driving force is modeled, similar to the experiments by \citet{Abel.2009}, which allow a systematic evaluation of the synchronization process.  Artificially created impulse trains are used as the input signal for the IPF. These so-called click tracks represent eighth notes played by a musician or a metronome. Nevertheless, all tempi in the following sections are given in $bpm$ based on quarter notes.

In most of the following Sections, it is helpful to focus on one tempo to maintain an overview. Therefore a tempo must be chosen, which feels most natural for most people. Several studies have been conducted to find this so-called \textit{spontaneous
motor tempo} (e.g., \cite{Collyer.1994,Parncutt.1994}) In the following, $120~bpm$ is chosen, which match the $2~Hz$ found by \citet{vanNoorden.1999}, as they have the same magnitude as the values given by most of the other studies.

All numerical simulations in this study are done using the programming language \textit{``Julia"} (see \citet{Bezanson.2017}).

\subsubsection{Discrete step changes}\label{sec:steps}

The first tempo changes which are systematically investigated are step changes, introduced by \citet{michon.1967}. Examples of the reaction of the IPF to those sudden, discrete changes in tempo are shown in Figure \ref{fig:Example}. The IPF and the click track always start at $120~bpm$, the  \textit{spontaneous motor tempo} introduced above. In general, three different reactions can occur:

\begin{itemize}
  \item The IPF converges to the tempo of the click track (see Figure \ref{fig:Example} a) and b))
  \item IPF is not effected by the click track, as the change in tempo was larger than a thirty-second note (see Figure \ref{fig:Example} c))
  \item IPF converges to a multiple tempo of the click track (double-time), due to the modulo operation described by Equation \eqref{eq:modulus} (see Figure \ref{fig:Example} d))
\end{itemize}

If the IPF synchronizes to the click track, it does not follow the changes instantaneously. Thus, there is always a transient oscillation. This is similar to humans reaction on sudden changes in tempo, where often a considerable \textit{overshoot} has been reported in literature (see e.g. \cite{michon.1967,Thaut.1998,Repp.2005,Repp.2001}). The IPF usually needs less than ten ticks until it synchronizes to the click track, similar to the cases discussed in literature. However, in contrast to those cases, the IPF shows more distinct oscillations. Listening to the audio examples provided by \cite{Linke.2021b}, this transition sounds familiar and natural, even though it may remind of a rather inexperienced musician.

It does not have to be a disadvantage that not all tempo fluctuations of the click track are followed by the IPF, as those fluctuations may refer to dotted notes or syncopations, and the IPF should follow the meter of the click track. In Figure \ref{fig:Example} c), the click track is slowed down from $120~bpm$ to $90~bpm$. Hence, a two against three rhythm occurs, as the IPF does not synchronize to the new tempo. Two against three rhythms can be produced using dotted notes or triplets, so this example shows that the IPF is not affected by those and remains to the fundamental meter.

\begin{figure}[htb]
  \begin{center}
  \includegraphics[width=.9\linewidth]{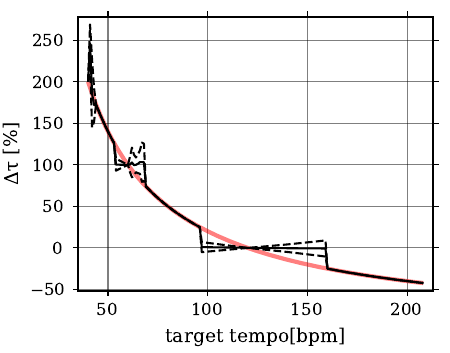} 
  \caption{Mean difference between the click track and the IPF $\Delta \tau$ (in percentage of the click track's tempo). The solid line refers to the difference, while the dashed lines refer to the standard derivation $\sigma$. The start tempo is always $120~bpm$, and the target tempo displayed at the axis of abscissas. The red line represents Equation \ref{eq:TheoDif}.}    
  \label{fig:StepDifference}
  \end{center}
\end{figure}

The steps size is now systematically varied: The click track always starts at $120~bpm$, and then steps to a different tempo in the range from $40$ to $208~bpm$ (resolution $1~bpm$), which equals the range of traditional metronomes (see \citet[p. 52]{vanNoorden.1999}). The click track has a total length of 32 eighth notes, where the tempo is changed after the first note. The results are shown in Figure \ref{fig:StepDifference}.

If the IPF would not synchronize to the click track, the difference $\Delta \tau$ between the tempos of click track and IPF could be described by the equation
\begin{equation}
   \widetilde{\Delta \tau}  = \frac{120-x}{x}~, \label{eq:TheoDif}
\end{equation}
where $x$ corresponds to the target tempo. This is in good accordance with most of the regions shown in Figure \ref{fig:StepDifference}. In these regions the standard derivation $\sigma$ of $\Delta \tau$ is negligible. As the click track and the IPF do not synchronize, their tempo does not change. Further, there is a salient region around $120~bpm$. Here, the IPF synchronizes to the click track, which results in $\Delta \tau=0$. $\sigma$ increases the more the target tempo differs from $120~bpm$. This corresponds to a stronger oscillation of the IPF when adapting to the new tempo. Similar but narrower regions can be found around $60~bpm$ and $30~bpm$, where $\Delta \tau$ is $100~\%$ resp. $200~\%$. As already described above, these regions synchronize to half- respectively quarter-time due to the modulo operation described by Equation \eqref{eq:modulus}.

\begin{figure}[htb]
  \begin{center}
  \includegraphics[width=.9 \linewidth]{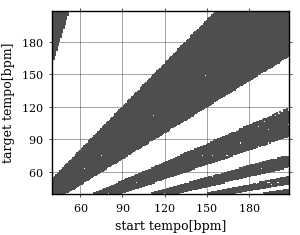} 
  \caption{Ability of the IPF to synchronize if the click track performs a step change from a tempo denoted at the abscissa to a tempo denoted at the ordinate. Gray regions correspond to synchronisation defined by $\sigma>10^{-12}$.}    
  \label{fig:StepHeat}
  \end{center}
\end{figure}

\begin{figure*}[htb]
  \begin{center}
  \includegraphics[width=.75 \linewidth]{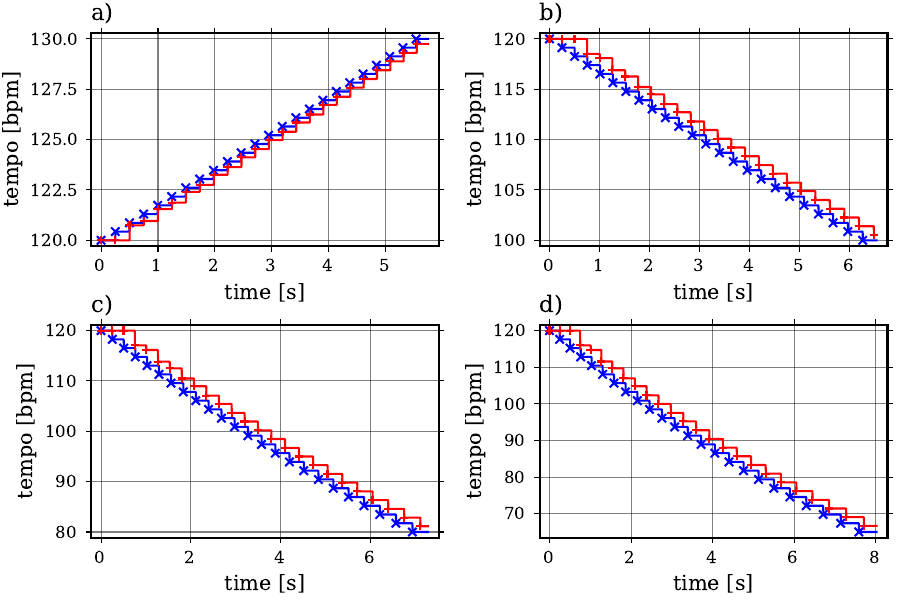} 
  \caption{The IPF reacting to the same changes in tempo as shown in Figure \ref{fig:Example}, when changing the tempo linear during 24 beats instead of step changes. Blue lines refer to the tempo of the click track. Red lines correspond to the tempo of the IPF. The single crosses represent single beats. Audio examples of these four cases can be found at \cite{Linke.2021b}}    
  \label{fig:Example_lin}
  \end{center}
\end{figure*}

It can be concluded that $\sigma>0$, indents synchronization. Therefore, every combination or start and target tempo is modeled to prove if this condition is fulfilled. To avoid numerical mistakes, only combinations where $\sigma > 10^{-12}$ are declared to synchronize. The results are shown in Figure \ref{fig:StepHeat}. Similar to Figure \ref{fig:StepDifference} the IPF synchronizes in a broad area around the initial tempo and some smaller regions related to the dividers provoked by Equation \ref{eq:modulus}. Further, $\sigma > 10^{-12}$ can be detected if the tempo is changed from very slow to swift tempi: This area shows the limitation of the synchronization threshold for changes larger than a thirty-second note. For instance, thirty-second notes at a quarter speed of $40~bpm$ could also be described as a sequence of $320~bpm$, which equals eighth notes at a quarter speed of $160~bpm$. Figure \ref{fig:StepHeat} shows that this is precisely the threshold for synchronization at large steps to swift tempi. 

\subsubsection{Linear changes}\label{sec:linear}

Now, instead of step changes, the tempo of the click track is changed linear based on the beats of the meter, not on time in seconds. Figure \ref{fig:Example_lin}  shows the same examples as Figure \ref{fig:Example}, but now the tempo change is applied linear during 24 beats, instead of suddenly from one beat two another. It is striking that here, the IPF is always capable of following the click track changes. Nevertheless, some minor phase differences $\Delta \Phi$ between the two signals are conspicuous.

\begin{figure}[htb]
  \begin{center}
  \includegraphics[width=.9 \linewidth]{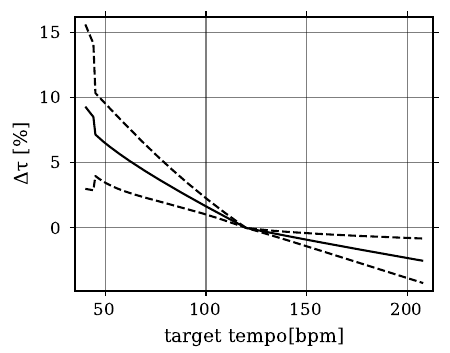} 
  \caption{Mean difference $\Delta \tau$ between the click track and the IPF (in the percentage of the Musicians tempo), while changing the tempo linear during 16 beats. The solid line refers to the difference, while the dashed lines refer to the standard derivation $\sigma$. The start tempo is always $120~bpm$, and the target tempo is displayed at the axis of abscissas. }    
  \label{fig:LinDifference}
  \end{center}
\end{figure}

Similar to Figure \ref{fig:StepDifference}, $\Delta \tau$ is systematically evaluated when changing the tempo linearly over 16 eighth notes in the range of a metronome ($40-208~bpm$). Again the transition always starts at $120~bpm$. Thus, different target tempos also refer to different slopes of the linear changes. The results are shown in Figure \ref{fig:LinDifference}. The overall $\Delta \tau$ are relative small, compared with Figure \ref{fig:StepDifference}. However, the absolute value of $\Delta \tau$  and $\sigma$ increase as the difference between start and target tempo, and thus the slope, is increased. As the IPF always reacts to the click track changes, its changes are usually too weak: When the click track increases the tempo, the IPF is too slow, and vice versa.

To evaluate the phase shift $\Delta \Phi$ between IPF and click track, which is visible in Figure \ref{fig:Example_lin}, the Pearson correlation coefficient $r$ is calculated for different lags. The IPF always reacts to the click track, so it must be approximately an eighth note too late. Thus, it must be sufficient to shift the IPF between zero seconds and a quarter note into the past, to find the maximum $r$. Thus, the lag at the maximum of $r$ determines $\Delta \Phi$.

\begin{figure}[htb]
  \begin{center}
  \includegraphics[width=.9 \linewidth]{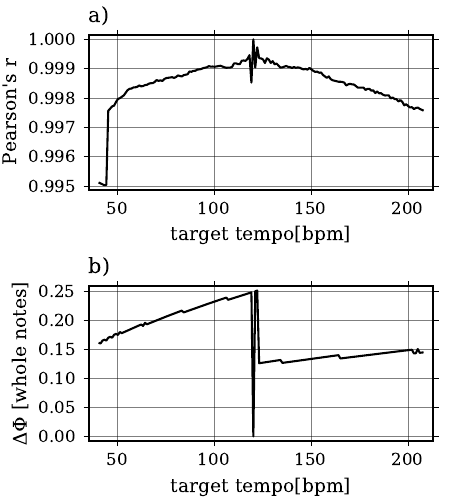} 
  \caption{a) Pearson's $r$ and b) $\Delta \Phi$, changing the tempo linear during 16 beats. $\Delta \Phi$ is given as the fraction of a whole note of the mean tempo of the click track. The start tempo is always $120~bpm$, and the target tempo displayed at the axis of abscissas.}    
  \label{fig:LinRs}
  \end{center}
\end{figure}

Figure \ref{fig:LinRs} a) shows the maximum $r$ for different target tempos. It is no surprise that the global maximum is at $120~bpm$. However, even though $r$ decreases if the slope of the tempo change increases, the overall values are surprisingly high. This is not a big surprise, as $\Delta \tau$ is rather small, and the Examples in Figure \ref{fig:Example_lin} are very similar when $\Delta \Phi$ is ignored.
Figure \ref{fig:LinRs} b) displays $\Delta \Phi$, given in whole notes of the mean tempo of the click track. Thus, as stated above, it must be approximately $0.125$ (an eighth note), in theory. However, this is only true if the tempo is slowly increased. A more significant slope increases $\Delta \Phi$. Remarkably the behavior is the other way around when the tempo is decreased. Slight decreases lead to high $\Delta \Phi$ up to a quarter note. Nevertheless, decreasing the tempo to lower tempos leads to small $\Delta \Phi$.

\subsubsection{Global perturbations}\label{sec:perturbations}
The examples discussed in Sections \ref{sec:steps} and \ref{sec:linear}, above are rather artificial. They assume the tempo of the click track to be constant and changes in tempo to happen suddenly or at least with a constant rate. When several Musicians are playing together in an ensemble, the tempo is perpetual fluctuating due to musical expression or the musicians' rhythmical skills (see, e.g., \cite{Repp.1992, Repp.2005}). The musicians' insufficiency can be simulated by adding different kinds of noise to the tempo of the click track, as discussed in the introduction.

\begin{figure*}[htb]
  \begin{center}
  \includegraphics[width=.75 \linewidth]{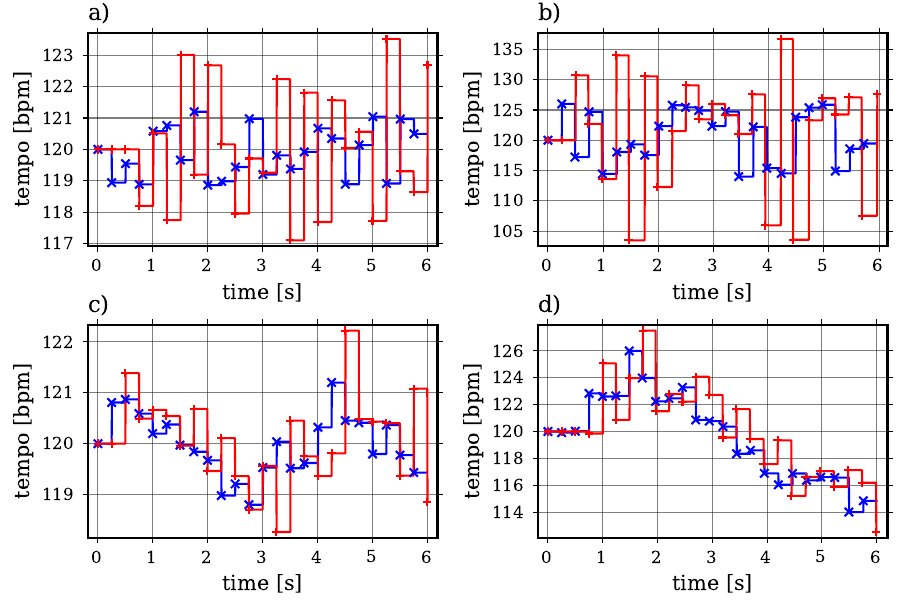} 
  \caption{IPF adapting to a noisy click track: the upper line (a) and b)) shows white noise and the lower line (c) and d)) Brownian noise. On the left (a) and c)), the fluctuation is $\pm 1~\%$, and on the right (b) and d)) $\pm 5~\%$. Blue lines refer to the tempo of the click track. Red lines correspond to the tempo of the IPF. The single crosses represent single beats. Audio examples of these four cases can be found at \cite{Linke.2021b}}    
  \label{fig:Example_noise}
  \end{center}
\end{figure*}

The examples presented in Figure \ref{fig:Example_noise} show different types (Brownian noise vs. white noise) and amounts ($1~\%$ and $5~\%$) of noise. A first observation shows that small amounts of noise and slow fluctuation lead to minor derivations between the IPF and the click track, while substantial and sudden fluctuations increase this error. Similar to the results shown in Figure \ref{fig:Example}, the IPS is likely to \textit{overshoot}, especially at large changes of tempo. A more systematic approach is made by calculating the same parameters as in Section \ref{sec:linear} (mean tempo difference $\Delta \tau$, phase difference $\Delta \Phi$ and Pearson's $r$) for different types and amounts of noise.
By comparing different studies \citet[p. 975]{Repp.2005} deduces three different groups of asynchronous behavior when participants tap to an isochronous metronome: Skilled percussionists show only $0.5~\%$ deviation, other musicians show approx. $2~\%$ and untrained participants show at least twice as large deviations. Thus, in the following, click tracks (32 eighth notes length) are created artificially, where either $0.5~\%$, $2~\%$ or $5~\%$ of noise are added to a constant tempo of $120~bpm$. As discussed in the introduction, most authors  (e.g., \cite{Coey.2016,Gilden.2001,vanOrden.2003}) state that pink noise is the best choice when modeling those minor imperfections. However, \citet{Hennig.2011} state that uncorrelated noise is still a standard solution in music technology and that sometimes $1/f^{1.3}$-noise would be more appropriated than $1/f$-noise. Therefore, this study focuses on the most common noise types: white, pink, or Brownian noise, respectively $1/f^0$-, $1/f^1$-, or $1/f^2$-noise. As the noise is stochastically generated for only 32 beats, the results may vary. Thus, each measurement is repeated ten times, and the mean values of $\Delta \tau$, $\Delta \Phi$, and $r$ are calculated.

\begin{table*}[htb]
  \centering
  \caption{Mean values of $\Delta \tau$, $\Delta \Phi$ and $r$ for different types and amounts of noise. The errors are the related $\sigma$. $\Delta \Phi$ is again given as the fraction of a whole note (4/4-note) of the mean tempo of the click track.}
  \label{tab:noise}
  \begin{tabular}{ l r| c c c}
    \hline \hline 
      
                                    &           & white noise       & pink noise        &Brownian noise     \\ \hline
    \multirow{3}{*}{mean $\Delta \tau$ [\%]}  & $0.5~\%$  & $0.02 \pm 0.69$   & $0.01 \pm 0.08$   & $0.01 \pm 0.24$   \\
                                    & $2~\%$    & $0.10 \pm 2.97$   & $0.08 \pm 1.58$   & $0.08 \pm 1.04$   \\
                                    & $5~\%$    & $0.31 \pm 6.65$   & $0.16 \pm 3.48$   & $0.24 \pm 3.15$   \\ \hline
    \multirow{3}{*}{mean $r$ }  & $0.5~\%$  & $0.58 \pm 0.12$   & $0.67 \pm 0.08$   & $0.85 \pm 0.08$   \\
                                    & $2~\%$    & $0.59 \pm 0.10$   & $0.68 \pm 0.12$   & $0.83 \pm 0.10$   \\
                                    & $5~\%$    & $0.56 \pm 0.09$   & $0.71 \pm 0.11$   & $0.76 \pm 0.12$  \\ \hline
    \multirow{3}{*}{mean $\Delta \Phi$ [4/4]} & $0.5~\%$  & $0.249 \pm 0.003$ & $0.237 \pm 0.038$ & $0.199 \pm 0.064$ \\
                                    & $2~\%$    & $0.245 \pm 0.003$ & $0.245 \pm 0.002$ & $0.162 \pm 0.059$ \\
                                    & $5~\%$    & $0.219 \pm 0.049$ & $0.207 \pm 0.057$ & $0.161 \pm 0.055$ \\ \hline \hline      
  \end{tabular} 
\end{table*}

The results shown in Table \ref{tab:noise} are similar to Figure \ref{fig:Example_noise}. The lower the amount of noise, and the higher the noise is correlated, the more precise the IPF can adapt to the click track. The type of noise seems to have a stronger impact than the amount. Compared to the linear changes shown in Figure \ref{fig:LinRs}, all $r$ are rather small. $\Delta \Phi$ is significantly higher than the theoretical value of $0.125$. Nevertheless, $\Delta \tau$ is small. Thus, this algorithm is still capable of real-time musical application as discussed in Section \ref{sec:AddBeta}.

As stated above, the tempo is more or less continuously modulated for musical expression during a musical performance. Compared to the minor unintended imperfections discussed above, those (relatively strong) modulations are perceived more distinctively and, according to \citet{Repp.2005} are more predictable. \citet{michon.1967} states that due to this predictability, it is possible to follow detectable and regular perturbations, like slow sinusoidal changes. This should also be true for the IPF, as it was already capable of following predictable linear changes in Section \ref{sec:linear}. 

\begin{figure*}[htb]
  \begin{center}
  \includegraphics[width=.75 \linewidth]{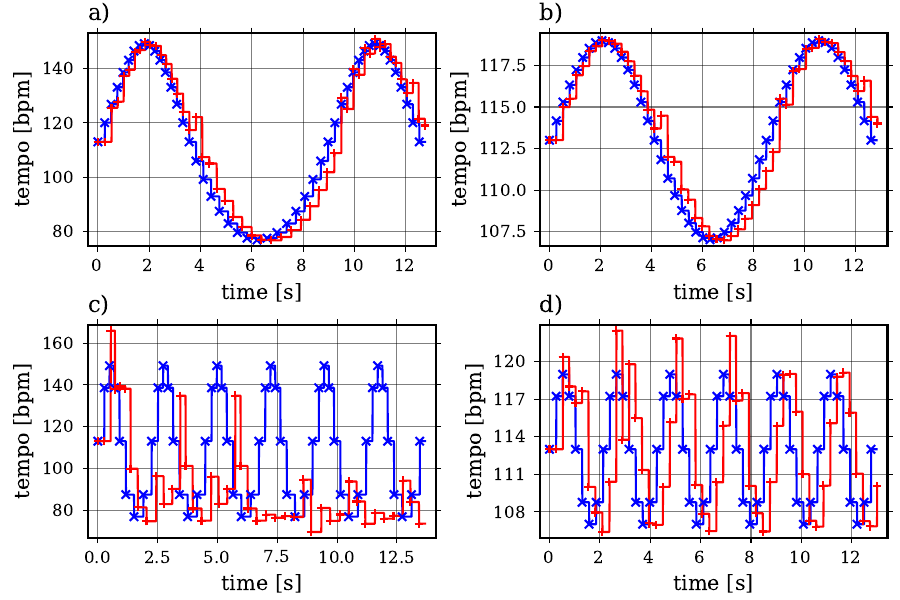} 
  \caption{IPF adapting to a sinusoidally modulated click track: the upper line (a) and b)) shows a modulation period of 32 eighth notes, and the lower line (c) and d)) shows a modulation period of 8 eighth notes. On the left (a) and c)), the amplitude is $36~bpm$, and on the right (b) and d)) $6~bpm$, both centered around $113~bpm$.  Blue lines refer to the tempo of the click track. Red lines correspond to the tempo of the IPF. The single crosses represent single beats. Audio examples of these four cases can be found at \cite{Linke.2021b}}    
  \label{fig:Example_sin}
  \end{center}
\end{figure*}

Examples of the IPf synchronizing to sinusoidal modulation are shown in Figure \ref{fig:Example_sin}. Slow modulation frequencies, shown in Figure \ref{fig:Example_sin} a) and b), can be followed by the IPF very reliable. It is remarkable that the fluctuation of the IPF's tempo in a) and b) looks very similar, even though the amplitude in a) is six times higher. In a) the overall shape of the sine is slightly distorted. This is due to the large amplitude, as the sine is defined relying on the beats and not time in seconds. 

As the frequency is increased in Figure \ref{fig:Example_sin} c) and d), the error of the IPF is increased, too. Further, the impact of the amplitude is increased. In c), large amplitudes, and thus changes in tempo, cannot be followed by the IPF anymore. The IPF seems to converge to a limit cycle close to the infimum of the click track. In d), the amplitude is significantly smaller. Here, the IPF can follow the tempo changes and adapts to the click track closer and closer. 

\begin{figure*}[htb]
  \begin{center}
  \includegraphics[width=1 \linewidth]{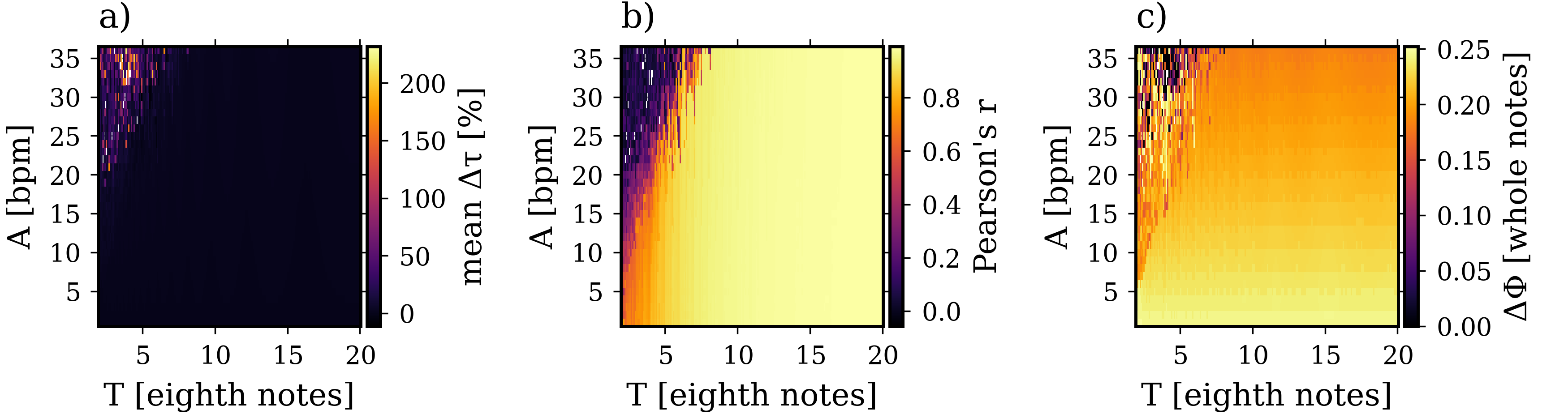} 
  \caption{a) mean $\Delta \tau$, b) $r$, and c) $\Delta \Phi$, when sinusoidal modulating the tempo around $113~bpm$ for different amplitudes and periodicities of the modulation function.}    
  \label{fig:CosHeat}
  \end{center}
\end{figure*}

For a comprehensive investigation of a sinusoidal click track, the amplitude $\hat{A}$ and the frequency $f$ must be varied systematically. A click track is created, which consists of 49 eighth notes. Then, considering the Nyquist–Shannon sampling theorem, the smallest period $T=1/f$ can be two eighth notes, and there is no limit for the longest $T$. According to \citet{vanNoorden.1999} the tempo should be varied between 75 and $150~bpm$ to stay in a range preferred by most people. This can be achieved by oscillating with an amplitude between zero and $36~bpm$ around $113~bpm$.

The results of the systematic parameter variation are shown in Figure \ref{fig:CosHeat}. For smaller periodicities, a clear proportionality can be observed: The longer $T$, the larger $\hat{A}$ can be chosen. $T$ larger than nine eighth notes do not show significant differences anymore. Here $\Delta \tau$ is negligible small, and $r>0.99$. The only differences which can be observed are for $\Delta \Phi$. It is always $>0.125$, but it decreases when $T$ and/or $\hat{A}$ are increased.

\subsection{Refining the IPF model}\label{sec:Refining} 
\subsubsection{Expanding the IPF model in respect to phase differences}\label{sec:AddBeta} 

Observing the shown examples of the IPF synchronizing to a click track (Figures \ref{fig:Example}, \ref{fig:Example_lin}, \ref{fig:Example_noise} and \ref{fig:Example_sin}) shows, that there not only a difference in tempo of those two signals but also in phase. In general, the beats of the click track do not take place at the same point in time as the beats of the IPF (blue x do not coincide with red +). Further, the measured $\Delta \Phi$ show that those phase differences are not a multiple of the click tracks tempo. This must not be a big issue. In real-time applications, small phase differences (latency) are unavoidable due to analog-to-digital conversion and the calculation time of the IPF (even if this is usually very fast). 

Further, phase differences always occur when synchronizing to an impulse train (even isochronous sequences), but they are corrected more easily and intuitively than tempo differences \cite{Keller.2007,Repp.2005,Repp.2001,Repp.2004}. Thus, they will not be a downside when using the IPF in a musical context, as here, not only the IPF adapts to a musician, but also a musician to the IPF. 

Nevertheless, those phase differences can be compensated by slightly extending the IPF: According to the linear time-keeper model by \citet{Wing.1973} the period $I_j$ between the $j^{th}$ beat of a musician (or the IPF) and the preceding beat can be described as 
\begin{equation}
  I_j=C_j-D_{j-1}+D_j ,
\end{equation}

where $C_j$ is the period of a precise internal clock. $D$ represents errors caused, e.g., by neuromuscular transmission lags or by movement time, where $D_j$ refers to the error of the present beat and $D_{j-1}$ of the last beat. Thus, the error of the present beat also depends on the error of the preceding beat. A performer realizes an error of a beat and corrects his next beat on this basis. This model has been applied to phase and periodicity synchronization in the past (e.g., \cite{Thaut.1998, Shaffer.1981}), and the translation to the IPF is straightforward. The IPF model derived in Section \ref{sec:IPFmodel} only describes the synchronization of the internal clock $C_j$ to an external time-keeper (e.g., another musician or a metronome). Adding another reflection point $\beta$ to the IPF represents the error of the motor system $D_j$. According to Equation \eqref{eq:IPF} the error of the preceding beat is inevitably considered due to the term $exp(g-g_-)$. Thus the extended formulation of the IPF is:

\begin{equation}
  g_+=g-ln \left( \frac 1 \alpha \left( g- \beta e^{g-g_{-}} \right) \right) , \label{eq:IPF_b}
\end{equation}

where $\alpha$ refers to the time between two beats of the click track and $\beta$ refers to the time difference between a beat of the click track and a beat of the IPF. According to \citet{Linke.2019b}, the IPF converges against $g=\alpha+\beta$ and thus $g$ represents the time between two beats of the IPF. If the tempo of the click track stays constant, $\beta$ approximates to zero (which annuls $\Delta \Phi$), and thus, $g$ approximates $\alpha$ (click track and IPF share the same pace).

\begin{figure}[htb]
  \begin{center}
    \includegraphics[width=.9 \linewidth]{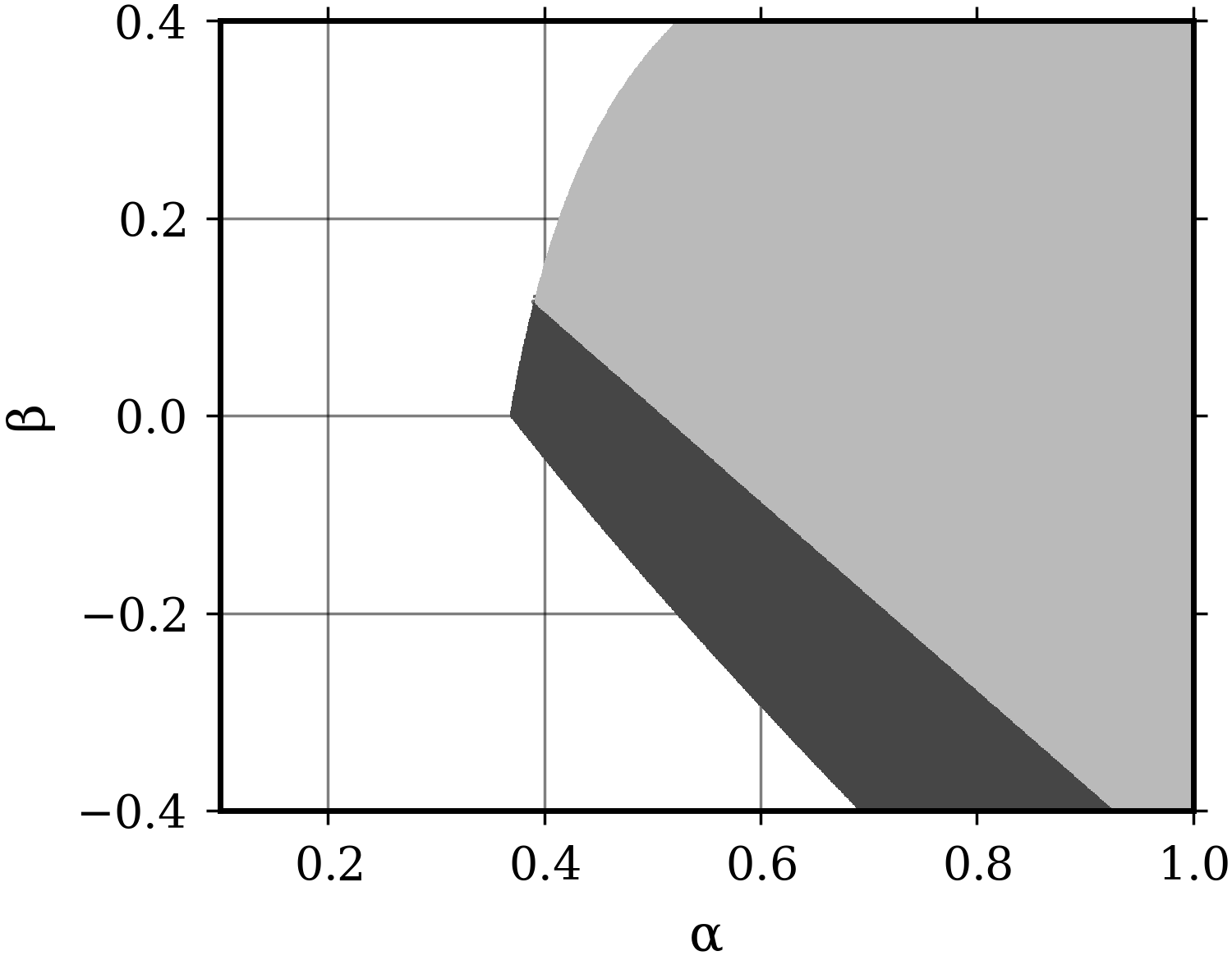} 
    \caption{Stability of the IPF with 2 reflection points in dependency of $\alpha$ and $\beta$. Gray regions are stable for every initial value $g_0 \in [g_s, 10]$, black regions never diverge $\forall g_0 \in [g_s, 10]$ but must not be stable, and white regions diverge at least for some $g_0 \in [g_s, 10]$, with the fixed point $g_s=\alpha+\beta$.}    
    \label{fig:stability_neg_beta}
  \end{center}
\end{figure}

\begin{figure*}[htb]
  \begin{center}
    \includegraphics[width=.75 \linewidth]{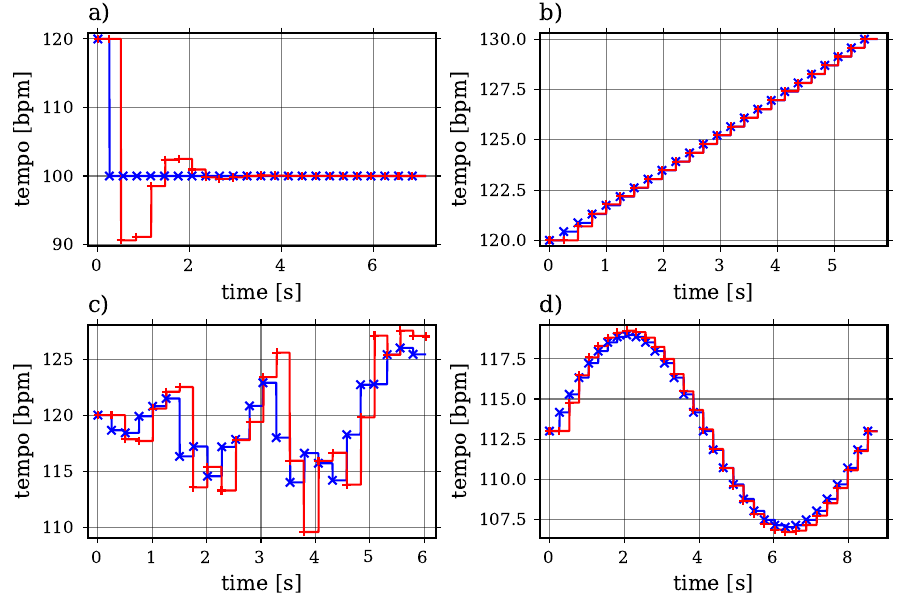} 
    \caption{Several scenarios shown in Figures \ref{fig:Example}, \ref{fig:Example_lin}, \ref{fig:Example_noise}, and \ref{fig:Example_sin} applied to the extended IPF of Equation \eqref{eq:IPF_b}: a) step change from 120 to $100~bpm$, b) linear change from 120 to $130~bpm$, c) $5~\%$ Brownian noise added to a $120~bpm$ click track and d) sinusoidal modulation with a period length of 32 eight notes varied $\pm 6~bpm$ around $113~bpm$.  Blue lines refer to the the tempo of the click track, red lines corresponds to the tempo of the IPF. The single crosses represent single beats. Audio examples of these four cases can be found at \cite{Linke.2021b}}    
    \label{fig:Example_beta}
  \end{center}
\end{figure*}

As $g$ is the sum of $\alpha$ and $\beta$ the scaling function should consist of linear terms only. Nevertheless, $\alpha_c$ should still relate to the maximum tempo $300~bpm$. This leads to a simplified version of Equation \eqref{eq:scaling_simple}:
\begin{equation}
  g,\alpha,\beta=5*\tau_{s} \ ,
\end{equation}

where $\tau_s$ relates to the period (in seconds) between two succeeding eighth notes. It is also worth noticing that here, $\beta$ may become negative - a case not covered in literature yet. Nevertheless, surprisingly, this does not affect the stability of the IPF when the absolute value of $\alpha$ remains sufficiently large, as shown in Figure \ref{fig:stability_neg_beta}. 

Examples of different applications of the extended Equation \eqref{eq:IPF_b} are shown in Figure \ref{fig:Example_beta}. Comparison with Figures \ref{fig:Example}, \ref{fig:Example_lin}, \ref{fig:Example_noise} and \ref{fig:Example_sin} shows an improvement for step changes (a)) where the transient seems to be somehow lowpass-filtered and afterwards all beats perfectly coincide. For linear and sinusoidal changes in tempo (b) and d)) the click track and the IPF are remarkably well-aligned. Even when noise is added (c)), the difference between IPF and click track seems to be reduced compared to Figure \ref{fig:Example_noise}. Nevertheless, the difference is not as striking as in the other cases. This is reasonable, as if the tempo is constantly changing, the phase relation changes, too. Thus there is no constant fixed point to which the IPF can converge.

\begin{table*}[htb]
  \centering
  \caption{Table \ref{tab:noise} recalculated using Equation \eqref{eq:IPF_b}}
  \label{tab:noise_beta}
  \begin{tabular}{ l r| c c c}
    \hline \hline 
      
                                    &           & white noise         & pink noise        &Brownian noise     \\ \hline
    \multirow{3}{*}{mean $\Delta \tau$ [\%]}  & $0.5~\%$  & $-0.001 \pm 0.581$  & $0.003 \pm 0.389$ & $0.003 \pm 0.231$ \\
                                    & $2~\%$    & $0.005 \pm 2.253$   & $0.008 \pm 1.330$ & $0.003 \pm 0.788$ \\
                                    & $5~\%$    & $0.10 \pm 6.13$     & $0.01 \pm 3.50$   & $-0.001 \pm 1.916$\\ \hline
    \multirow{3}{*}{mean $r$ }      & $0.5~\%$  & $0.97 \pm 0.01$     & $0.95 \pm 0.01$   & $0.94 \pm 0.01$   \\
                                    & $2~\%$    & $0.96 \pm 0.01$     & $0.95 \pm 0.01$   & $0.95 \pm 0.01$   \\
                                    & $5~\%$    & $0.95 \pm 0.01$     & $0.95 \pm 0.01$   & $0.95 \pm 0.01$   \\ \hline
   \multirow{3}{*}{mean $\Delta \Phi$ [4/4]} & $0.5~\%$  & $0.125 \pm 0.001$   & $0.125 \pm 0.001$ & $0.125 \pm 0.001$ \\
                                    & $2~\%$    & $0.125 \pm 0.001$   & $0.125 \pm 0.001$ & $0.125 \pm 0.001$ \\
                                    & $5~\%$    & $0.125 \pm 0.001$   & $0.124 \pm 0.001$ & $0.125 \pm 0.002$ \\ \hline \hline      
  \end{tabular} 
\end{table*}

A first in-depth comparison is made, focussing on the perturbations caused by noise. Therefore the values of Table \ref{tab:noise} are recalculated using Equation \eqref{eq:IPF_b}. The results are shown in Table \ref{tab:noise_beta}. All values have been significantly improved: The correlations are higher (all $\approx 0.95$), the tempo differences lower, and all $\sigma$ are lower, too. No significant difference between the different types of noise can be recognized anymore. Sometimes even white noise shows better synchronization than the other types. Remarkably, nearly all $\Delta \Phi=1/8$, which is exactly the value we would expect, as the IPF always reacts an eighth note too late. Thus, the phases correction of the IPF seems to work, even if the tempo is not constant. 

\subsubsection{Polyrhythmic tempo detection}\label{sec:center120bpm} 

In some musical applications, it might be wanted that the IPF synchronizes to a broader range of tempos (not like shown in Figure \ref{fig:StepDifference}) or that the bandwidth of possible tempos for the IPF is reduced to prevent large tempo changes. Both can be done by allowing polyrhythmic beats.

\citet{Handel.1981} investigated how people tap along polyrhythmic patterns. Applying models of damped resonating oscillators on their data \citet{vanNoorden.1999} conclude that most people prefer tapping to sub-patterns with a frequency close to $120~bpm$. The approach can be inverted and transferred to tempo detection of the IPF model. Not the actual tempo of the click track is taken into account, but a different tempo closer to $120~bpm$, which builds up a regular polyrhythm to the click track. Therefore only minor changes need to be applied to the tempo detection method introduced in Section \ref{sec:IPFmodel}:
\begin{itemize}
  \item First, appropriated polyrhythmic relations have to be chosen. Here, the ratios used by \citet{Handel.1981} (2/3, 3/4, 2/5, 3/5, and 4/5) as well as their non-polyrhythmic counterparts (1/1, 1/2, 1/3, 1/4, and 1/5) are chosen, which corresponds to a Farey sequence of order $5$ (see, e.g., \citet{Haken.1996}, \citet[pp. 515]{Argyris.2015}). 
  \item In Section \ref{sec:IPFmodel} the tempo was investigated by calculating the time difference between two consecutive beats of the click track. Now, this difference is multiplied by each of the factors chosen above as well as their reciprocal. 
  \item The result closest to $250~ms$ (which relates to eighth notes at $120~bpm$) is chosen as the new tempo.
  \item This way, differences larger than a thirty-second note must no longer be ignored.
\end{itemize}

\begin{figure*}[htb]
  \begin{center}
    \includegraphics[width=.75 \linewidth]{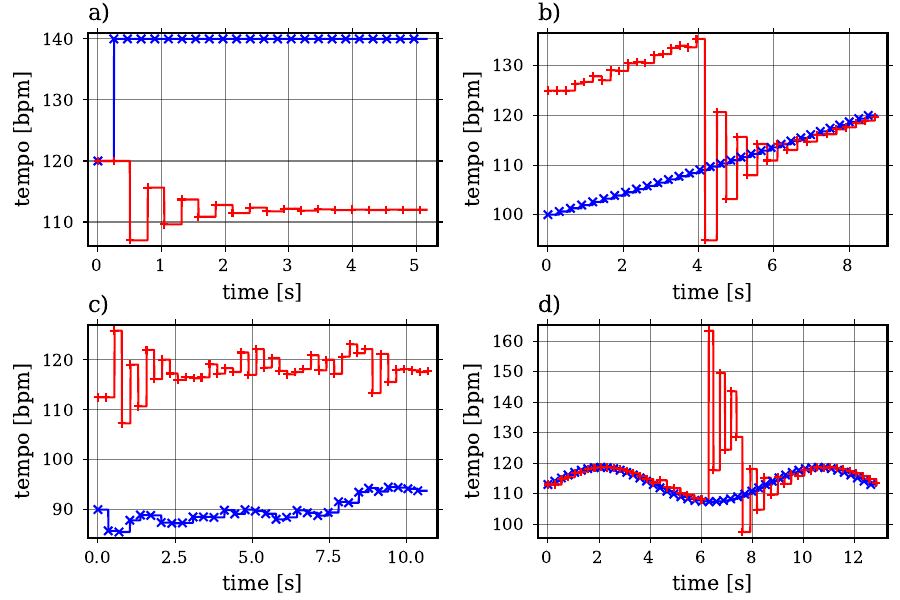} 
    \caption{Several scenarios shown in Figures \ref{fig:Example}, \ref{fig:Example_lin}, \ref{fig:Example_noise}, and \ref{fig:Example_sin} applied to polyrhythmic tempo detection: a) step change from 120 to $140~bpm$, b) linear change from 100 to $120~bpm$, c) $5~\%$ Brownian noise added to a $90~bpm$ click track and d) sinusoidal modulation with a period length of 32 eight notes varied $\pm 6~bpm$ around $113~bpm$.  Blue lines refer to the the tempo of the click track, red lines corresponds to the tempo of the IPF. The single crosses represent single beats. Audio examples of these four cases can be found at \cite{Linke.2021b}}    
    \label{fig:Example_poly}
  \end{center}
\end{figure*}

Combining this tempo detection with the most simple IPF model derived in Section \ref{sec:IPFmodel} and determined by Equation \eqref{eq:IPF_simple}, the IPF is capable of synchronizing in a polyrhythmic manner. Examples of different applications of the polyrhythmic tempo detection are shown in Figure \ref{fig:Example_poly}. All Examples are related to four against five polyrhythms. In a) the click track suddenly jumps from $120~bpm$ to $140~bpm$. Instead of following this increase (like the algorithm in Section \ref{sec:IPFBehavior} would have done), the IPF slows down to $112~bpm$ resulting in a $4/5$ ratio. In b), the tempo of the click track is increased linearly from $100~bpm$ to $120~bpm$. Here the IPF starts with a $4/5$ ratio at $125~bpm$ and follows the increase of the click track with the same ratio. However, in contrast to Figure \ref{fig:Example_lin}, the linear increase is a bit noisier. The prediction of the click tracks tempo changes does not work as well as in Figure \ref{fig:Example_lin}, either. As soon as the click track reaches $110~bpm$, the IPf starts to synchronize in a 1/1 ratio. Similar results can be seen in d). The IPF can still follow the sinusoidal fluctuations around $130~bpm$, but as soon as the click track drops below $110~bpm$, the IPF tries to synchronize with a $4/5$ ratio. As the settling time of the IPF is too long, the click track is above $110~bpm$ again, before the IPF has reached a $4/5$ ratio and the IPF synchronizes at a 1/1 ratio again. Finally, in c), the click track fluctuates around $90~bpm$, and the IPF follows with a $4/5$ ratio around $112.5~bpm$. The IPF can follow perturbations caused by Brownian noise, but it seems to overreact in some cases.

\begin{figure}[htb]
  \begin{center}
  \includegraphics[width=.9 \linewidth]{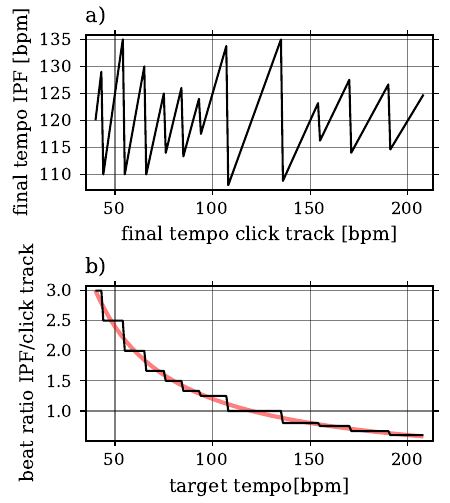} 
  \caption{IPFs behavior on different step functions. The start tempo is always $120~bpm$, and the final tempo $T_F$ of the click track (32 eighth notes long) is always displayed at the axis of abscissas. a) shows the corresponding final tempo of the IPF and b) shows the related ratio of final tempos of IPF over click track (which corresponds to the different polyrhythmic ratios). The red line represents $120/T_F$, which is true if no synchronization happens.}    
  \label{fig:PolyDifference}
  \end{center}
\end{figure}

The behavior of this tempo detection can be systematically investigated by applying different step changes in tempo, similar to Section \ref{sec:steps}. The results are shown in Figure \ref{fig:PolyDifference}. Focussing on b) one can see the different polyrhythmic ratios as discrete steps in the ratio of the different final tempos. The red line indicates that the click track and the IPF always synchronize. Further, a) shows that now the tempo range of the IPF is significantly reduced to $108-135~bpm$ and sudden steps in tempo occur when the polyrhythmic ratio is changed. The bandwidth of the IPF could be extended again if the number of possible polyrhythmic ratios is reduced. Further, the favorable $120~bpm$ in the tempo detection can be easily changed to any arbitrary value to shift the rhythmical center, if this is helpful for specific musical applications.

\section{Discussion}\label{sec:Conclusion} 

Applying the IPF on the rhythmical synchronization of musicians is a straightforward modeling approach that leads to convincing results, even in the most simple form defined by Equation \eqref{eq:IPF_simple}. Similar to humans, the IPF adapts to step changes in tempo after a short, chaotic transition and is capable of predicting regular changes in tempo and quickly adapts to them, like discussed in Sections \ref{sec:linear} and \ref{sec:perturbations}. Applied to noisy signals, the IPF shows a mixture of those two behaviors. However, the required algorithm for tempo detection and the scaling of the system parameters can have a crucial impact on the results gained by the IPF.

It could be shown that the number of details in the simulation process and the related results can be easily adjusted by varying the number of reflection points $\beta_k$. Even though the results are convincing when using Equation \eqref{eq:IPF_simple}, the improvements when adding $\beta$ in Section \ref{sec:AddBeta} are striking. Of course, in future work, this approach can be extended by adding further $\beta_k$. Here, two different strategies are conceivable. Similar to Section \ref{sec:AddBeta} further $\beta_k$ could represent the physiology of a musician, e.g., electrical impulses in the muscular system, the cochlear nerve, or the motor cortex. Alternatively, further $\beta_k$ could represent further musicians, e.g., when synchronizing to a chamber music ensemble or a symphony orchestra with its maestro. The only differences between these two strategies are the strengths and possible time dependencies of the single $\beta_k$.

As described in Section \ref{sec:AddBeta} more precise IPF models may not necessarily be an advantage for real-time applications. System latencies, such as analog-to-digital conversions, are usually unavoidable, and thus $\Delta \Phi$ cannot entirely be corrected. An alternative approach to increasing the IPF model's performance is to consider theoretical findings, e.g., from music psychology. The utilization of this strategy has been shown in Section \ref{sec:center120bpm}. However, Section \ref{sec:center120bpm} also shows that adding more theoretical assumptions restricts the free, chaotic, and self-organizing behavior of the IPF, which usually is one of the compelling advantages of this modeling technic. Thus the restrictions must be carefully balanced.

In this study, the IPF was judged by comparing it with the results of different tapping tests. According to \citet{Pikovsky.2001}, this a not a proper synchronization process, as the artificially generated click track does not respond to the IPF. Thus, in future research, the coupling of the IPF to a human musician should be investigated, like the coupling between musicians, which, compared to tapping experiments, has not been investigated extensively in the past (see, e.g., \cite{Keller.2007,Sogorski.2018,Hennig.2014,Muller.2013}). Those studies should result in more complex synchronization scenarios.

Besides a formal description and an in-depth analysis of the rhythmical synchronization process of a musician, the proposed IPF model can be used to replace drum machines and click tracks with more musical and creative solutions in live music and recording scenarios. By fine-tuning the modeling parameters, one can decide whether to model the realistic behavior of a human musician or to create novel, unfamiliar systems with a distinct, characteristic manner of rhythmical fluctuation and synchronization. Further, in future work, identification of drummers due to individual timing might also be possible using the presented system. 

\bibliography{DrumSync_bib.bib}

\begin{thebibliography}{}

\bibitem [\protect \citeauthoryear {%
Abel%
, Ahnert%
\BCBL {}\ \BBA {} Bergweiler%
}{%
Abel%
\ \protect \BOthers {.}}{%
{\protect \APACyear {2009}}%
}]{%
Abel.2009}
\APACinsertmetastar {%
Abel.2009}%
\begin{APACrefauthors}%
Abel, M.%
, Ahnert, K.%
\BCBL {}\ \BBA {} Bergweiler, S.%
\end{APACrefauthors}%
\unskip\
\newblock
\APACrefYearMonthDay{2009}{}{}.
\newblock
{\BBOQ}\APACrefatitle {Synchronization of sound sources} {Synchronization of
  sound sources}.{\BBCQ}
\newblock
\APACjournalVolNumPages{Physical review letters}{103}{11}{114301}.
\newblock
\begin{APACrefDOI} \doi{10.1103/PhysRevLett.103.114301} \end{APACrefDOI}
\PrintBackRefs{\CurrentBib}

\bibitem [\protect \citeauthoryear {%
Argyris%
, Faust%
, Haase%
\BCBL {}\ \BBA {} Friedrich%
}{%
Argyris%
\ \protect \BOthers {.}}{%
{\protect \APACyear {2015}}%
}]{%
Argyris.2015}
\APACinsertmetastar {%
Argyris.2015}%
\begin{APACrefauthors}%
Argyris, J\BPBI H.%
, Faust, G.%
, Haase, M.%
\BCBL {}\ \BBA {} Friedrich, R.%
\end{APACrefauthors}%
\unskip\
\newblock
\APACrefYear{2015}.
\newblock
\APACrefbtitle {An Exploration of Dynamical Systems and Chaos: Completely
  Revised and Enlarged Second Edition} {An exploration of dynamical systems and
  chaos: Completely revised and enlarged second edition}.
\newblock
\APACaddressPublisher{Berlin, Heidelberg}{{Springer Berlin Heidelberg}}.
\newblock
\begin{APACrefDOI} \doi{10.1007/978-3-662-46042-9} \end{APACrefDOI}
\PrintBackRefs{\CurrentBib}

\bibitem [\protect \citeauthoryear {%
Bader%
}{%
Bader%
}{%
{\protect \APACyear {2013}}%
}]{%
Bader.2013}
\APACinsertmetastar {%
Bader.2013}%
\begin{APACrefauthors}%
Bader, R.%
\end{APACrefauthors}%
\unskip\
\newblock
\APACrefYear{2013}.
\newblock
\APACrefbtitle {Nonlinearities and synchronization in musical acoustics and
  music psychology} {Nonlinearities and synchronization in musical acoustics
  and music psychology}\ (\BVOL~2).
\newblock
\APACaddressPublisher{Berlin [et al.]}{Springer}.
\newblock
\begin{APACrefDOI} \doi{10.1007/978-3-642-36098-5} \end{APACrefDOI}
\PrintBackRefs{\CurrentBib}

\bibitem [\protect \citeauthoryear {%
Bezanson%
, Edelman%
, Karpinski%
\BCBL {}\ \BBA {} Shah%
}{%
Bezanson%
\ \protect \BOthers {.}}{%
{\protect \APACyear {2017}}%
}]{%
Bezanson.2017}
\APACinsertmetastar {%
Bezanson.2017}%
\begin{APACrefauthors}%
Bezanson, J.%
, Edelman, A.%
, Karpinski, S.%
\BCBL {}\ \BBA {} Shah, V\BPBI B.%
\end{APACrefauthors}%
\unskip\
\newblock
\APACrefYearMonthDay{2017}{}{}.
\newblock
{\BBOQ}\APACrefatitle {Julia: A Fresh Approach to Numerical Computing} {Julia:
  A fresh approach to numerical computing}.{\BBCQ}
\newblock
\APACjournalVolNumPages{SIAM Review}{59}{1}{65--98}.
\newblock
\begin{APACrefDOI} \doi{10.1137/141000671} \end{APACrefDOI}
\PrintBackRefs{\CurrentBib}

\bibitem [\protect \citeauthoryear {%
Coey%
, Washburn%
, Hassebrock%
\BCBL {}\ \BBA {} Richardson%
}{%
Coey%
\ \protect \BOthers {.}}{%
{\protect \APACyear {2016}}%
}]{%
Coey.2016}
\APACinsertmetastar {%
Coey.2016}%
\begin{APACrefauthors}%
Coey, C\BPBI A.%
, Washburn, A.%
, Hassebrock, J.%
\BCBL {}\ \BBA {} Richardson, M\BPBI J.%
\end{APACrefauthors}%
\unskip\
\newblock
\APACrefYearMonthDay{2016}{}{}.
\newblock
{\BBOQ}\APACrefatitle {Complexity matching effects in bimanual and
  interpersonal syncopated finger tapping} {Complexity matching effects in
  bimanual and interpersonal syncopated finger tapping}.{\BBCQ}
\newblock
\APACjournalVolNumPages{Neuroscience letters}{616}{}{204--210}.
\newblock
\begin{APACrefDOI} \doi{10.1016/j.neulet.2016.01.066} \end{APACrefDOI}
\PrintBackRefs{\CurrentBib}

\bibitem [\protect \citeauthoryear {%
Collyer%
, Broadbent%
\BCBL {}\ \BBA {} Church%
}{%
Collyer%
\ \protect \BOthers {.}}{%
{\protect \APACyear {1994}}%
}]{%
Collyer.1994}
\APACinsertmetastar {%
Collyer.1994}%
\begin{APACrefauthors}%
Collyer, C\BPBI E.%
, Broadbent, H\BPBI A.%
\BCBL {}\ \BBA {} Church, R\BPBI M.%
\end{APACrefauthors}%
\unskip\
\newblock
\APACrefYearMonthDay{1994}{}{}.
\newblock
{\BBOQ}\APACrefatitle {Preferred rates of repetitive tapping and categorical
  time production} {Preferred rates of repetitive tapping and categorical time
  production}.{\BBCQ}
\newblock
\APACjournalVolNumPages{Perception {\&} Psychophysics}{55}{4}{443--453}.
\newblock
\begin{APACrefDOI} \doi{10.3758/bf03205301} \end{APACrefDOI}
\PrintBackRefs{\CurrentBib}

\bibitem [\protect \citeauthoryear {%
Datseris%
\ \protect \BOthers {.}}{%
Datseris%
\ \protect \BOthers {.}}{%
{\protect \APACyear {2019}}%
}]{%
Datseris.2019}
\APACinsertmetastar {%
Datseris.2019}%
\begin{APACrefauthors}%
Datseris, G.%
, Ziereis, A.%
, Albrecht, T.%
, Hagmayer, Y.%
, Priesemann, V.%
\BCBL {}\ \BBA {} Geisel, T.%
\end{APACrefauthors}%
\unskip\
\newblock
\APACrefYearMonthDay{2019}{}{}.
\newblock
{\BBOQ}\APACrefatitle {Microtiming Deviations and Swing Feel in Jazz}
  {Microtiming deviations and swing feel in jazz}.{\BBCQ}
\newblock
\APACjournalVolNumPages{Scientific reports}{9}{1}{19824}.
\newblock
\begin{APACrefDOI} \doi{10.1038/s41598-019-55981-3} \end{APACrefDOI}
\PrintBackRefs{\CurrentBib}

\bibitem [\protect \citeauthoryear {%
D'Huys%
, J{\"u}ngling%
\BCBL {}\ \BBA {} Kinzel%
}{%
D'Huys%
\ \protect \BOthers {.}}{%
{\protect \APACyear {2016}}%
}]{%
DHuys.2016}
\APACinsertmetastar {%
DHuys.2016}%
\begin{APACrefauthors}%
D'Huys, O.%
, J{\"u}ngling, T.%
\BCBL {}\ \BBA {} Kinzel, W.%
\end{APACrefauthors}%
\unskip\
\newblock
\APACrefYearMonthDay{2016}{}{}.
\newblock
{\BBOQ}\APACrefatitle {On the Interplay of Noise and Delay in Coupled
  Oscillators} {On the interplay of noise and delay in coupled
  oscillators}.{\BBCQ}
\newblock
\BIn{} \APACrefbtitle {Control of Self-Organizing Nonlinear Systems} {Control
  of self-organizing nonlinear systems}\ (\BPGS\ 127--145).
\newblock
\APACaddressPublisher{}{{Springer, Cham}}.
\newblock
\begin{APACrefURL}
  \url{https://link.springer.com/chapter/10.1007/978-3-319-28028-8_7}
  \end{APACrefURL}
\newblock
\begin{APACrefDOI} \doi{10.1007/978-3-319-28028-8_7} \end{APACrefDOI}
\PrintBackRefs{\CurrentBib}

\bibitem [\protect \citeauthoryear {%
Fletcher%
\ \BBA {} Rossing%
}{%
Fletcher%
\ \BBA {} Rossing%
}{%
{\protect \APACyear {2010}}%
}]{%
Fletcher.2010}
\APACinsertmetastar {%
Fletcher.2010}%
\begin{APACrefauthors}%
Fletcher, N\BPBI H.%
\BCBT {}\ \BBA {} Rossing, T\BPBI D.%
\end{APACrefauthors}%
\unskip\
\newblock
\APACrefYear{2010}.
\newblock
\APACrefbtitle {The physics of musical instruments} {The physics of musical
  instruments}\ (\PrintOrdinal{2. ed., [rpt..]}\ \BEd).
\newblock
\APACaddressPublisher{New York, NY}{Springer}.
\newblock
\begin{APACrefDOI} \doi{10.1007/978-0-387-21603-4} \end{APACrefDOI}
\PrintBackRefs{\CurrentBib}

\bibitem [\protect \citeauthoryear {%
Gilden%
}{%
Gilden%
}{%
{\protect \APACyear {2001}}%
}]{%
Gilden.2001}
\APACinsertmetastar {%
Gilden.2001}%
\begin{APACrefauthors}%
Gilden, D\BPBI L.%
\end{APACrefauthors}%
\unskip\
\newblock
\APACrefYearMonthDay{2001}{}{}.
\newblock
{\BBOQ}\APACrefatitle {Cognitive emissions of 1/f noise} {Cognitive emissions
  of 1/f noise}.{\BBCQ}
\newblock
\APACjournalVolNumPages{Psychological review}{108}{1}{33--56}.
\newblock
\begin{APACrefDOI} \doi{10.1037/0033-295x.108.1.33} \end{APACrefDOI}
\PrintBackRefs{\CurrentBib}

\bibitem [\protect \citeauthoryear {%
Haken%
, Kelso%
\BCBL {}\ \BBA {} Bunz%
}{%
Haken%
\ \protect \BOthers {.}}{%
{\protect \APACyear {1985}}%
}]{%
Haken.1985}
\APACinsertmetastar {%
Haken.1985}%
\begin{APACrefauthors}%
Haken, H.%
, Kelso, J\BPBI A.%
\BCBL {}\ \BBA {} Bunz, H.%
\end{APACrefauthors}%
\unskip\
\newblock
\APACrefYearMonthDay{1985}{}{}.
\newblock
{\BBOQ}\APACrefatitle {A theoretical model of phase transitions in human hand
  movements} {A theoretical model of phase transitions in human hand
  movements}.{\BBCQ}
\newblock
\APACjournalVolNumPages{Biological cybernetics}{51}{5}{347--356}.
\newblock
\begin{APACrefDOI} \doi{10.1007/BF00336922} \end{APACrefDOI}
\PrintBackRefs{\CurrentBib}

\bibitem [\protect \citeauthoryear {%
Haken%
, Peper%
, Beek%
\BCBL {}\ \BBA {} Daffertshofer%
}{%
Haken%
\ \protect \BOthers {.}}{%
{\protect \APACyear {1996}}%
}]{%
Haken.1996}
\APACinsertmetastar {%
Haken.1996}%
\begin{APACrefauthors}%
Haken, H.%
, Peper, C\BPBI E.%
, Beek, P\BPBI J.%
\BCBL {}\ \BBA {} Daffertshofer, A.%
\end{APACrefauthors}%
\unskip\
\newblock
\APACrefYearMonthDay{1996}{}{}.
\newblock
{\BBOQ}\APACrefatitle {A model for phase transitions in human hand movements
  during multifrequency tapping} {A model for phase transitions in human hand
  movements during multifrequency tapping}.{\BBCQ}
\newblock
\APACjournalVolNumPages{Physica D: Nonlinear Phenomena}{90}{1-2}{179--196}.
\newblock
\begin{APACrefURL}
  \url{https://www.sciencedirect.com/science/article/pii/0167278995002359}
  \end{APACrefURL}
\newblock
\begin{APACrefDOI} \doi{10.1016/0167-2789(95)00235-9} \end{APACrefDOI}
\PrintBackRefs{\CurrentBib}

\bibitem [\protect \citeauthoryear {%
Handel%
\ \BBA {} Oshinsky%
}{%
Handel%
\ \BBA {} Oshinsky%
}{%
{\protect \APACyear {1981}}%
}]{%
Handel.1981}
\APACinsertmetastar {%
Handel.1981}%
\begin{APACrefauthors}%
Handel, S.%
\BCBT {}\ \BBA {} Oshinsky, J\BPBI S.%
\end{APACrefauthors}%
\unskip\
\newblock
\APACrefYearMonthDay{1981}{}{}.
\newblock
{\BBOQ}\APACrefatitle {The meter of syncopated auditory polyrhythms} {The meter
  of syncopated auditory polyrhythms}.{\BBCQ}
\newblock
\APACjournalVolNumPages{Perception {\&} Psychophysics}{30}{1}{1--9}.
\PrintBackRefs{\CurrentBib}

\bibitem [\protect \citeauthoryear {%
Hennig%
}{%
Hennig%
}{%
{\protect \APACyear {2014}}%
}]{%
Hennig.2014}
\APACinsertmetastar {%
Hennig.2014}%
\begin{APACrefauthors}%
Hennig, H.%
\end{APACrefauthors}%
\unskip\
\newblock
\APACrefYearMonthDay{2014}{}{}.
\newblock
{\BBOQ}\APACrefatitle {Synchronization in human musical rhythms and mutually
  interacting complex systems} {Synchronization in human musical rhythms and
  mutually interacting complex systems}.{\BBCQ}
\newblock
\APACjournalVolNumPages{Proceedings of the National Academy of Sciences of the
  United States of America}{111}{36}{12974--12979}.
\newblock
\begin{APACrefDOI} \doi{10.1073/pnas.1324142111} \end{APACrefDOI}
\PrintBackRefs{\CurrentBib}

\bibitem [\protect \citeauthoryear {%
Hennig%
\ \protect \BOthers {.}}{%
Hennig%
\ \protect \BOthers {.}}{%
{\protect \APACyear {2011}}%
}]{%
Hennig.2011}
\APACinsertmetastar {%
Hennig.2011}%
\begin{APACrefauthors}%
Hennig, H.%
, Fleischmann, R.%
, Fredebohm, A.%
, Hagmayer, Y.%
, Nagler, J.%
, Witt, A.%
\BDBL {}Geisel, T.%
\end{APACrefauthors}%
\unskip\
\newblock
\APACrefYearMonthDay{2011}{}{}.
\newblock
{\BBOQ}\APACrefatitle {The nature and perception of fluctuations in human
  musical rhythms} {The nature and perception of fluctuations in human musical
  rhythms}.{\BBCQ}
\newblock
\APACjournalVolNumPages{PloS one}{6}{10}{e26457}.
\newblock
\begin{APACrefDOI} \doi{10.1371/journal.pone.0026457} \end{APACrefDOI}
\PrintBackRefs{\CurrentBib}

\bibitem [\protect \citeauthoryear {%
Hennig%
, Fleischmann%
\BCBL {}\ \BBA {} Geisel%
}{%
Hennig%
\ \protect \BOthers {.}}{%
{\protect \APACyear {2012}}%
}]{%
Hennig.2012}
\APACinsertmetastar {%
Hennig.2012}%
\begin{APACrefauthors}%
Hennig, H.%
, Fleischmann, R.%
\BCBL {}\ \BBA {} Geisel, T.%
\end{APACrefauthors}%
\unskip\
\newblock
\APACrefYearMonthDay{2012}{}{}.
\newblock
{\BBOQ}\APACrefatitle {Musical rhythms: The science of being slightly off}
  {Musical rhythms: The science of being slightly off}.{\BBCQ}
\newblock
\APACjournalVolNumPages{Physics Today}{65}{7}{64--65}.
\newblock
\begin{APACrefURL}
  \url{https://www.researchgate.net/profile/Holger-Hennig/publication/257020003_Musical_rhythms_The_science_of_being_slightly_off/links/56b3a0f408ae1f8aa45350dc/Musical-rhythms-The-science-of-being-slightly-off.pdf}
  \end{APACrefURL}
\newblock
\begin{APACrefDOI} \doi{10.1063/PT.3.1650} \end{APACrefDOI}
\PrintBackRefs{\CurrentBib}

\bibitem [\protect \citeauthoryear {%
Just%
, Geffert%
, Zakharova%
\BCBL {}\ \BBA {} Sch{\"o}ll%
}{%
Just%
\ \protect \BOthers {.}}{%
{\protect \APACyear {2016}}%
}]{%
Just.2016}
\APACinsertmetastar {%
Just.2016}%
\begin{APACrefauthors}%
Just, W.%
, Geffert, P\BPBI M.%
, Zakharova, A.%
\BCBL {}\ \BBA {} Sch{\"o}ll, E.%
\end{APACrefauthors}%
\unskip\
\newblock
\APACrefYearMonthDay{2016}{}{}.
\newblock
{\BBOQ}\APACrefatitle {Noisy Dynamical Systems with Time Delay: Some Basic
  Analytical Perturbation Schemes with Applications} {Noisy dynamical systems
  with time delay: Some basic analytical perturbation schemes with
  applications}.{\BBCQ}
\newblock
\BIn{} \APACrefbtitle {Control of Self-Organizing Nonlinear Systems} {Control
  of self-organizing nonlinear systems}\ (\BPGS\ 147--168).
\newblock
\APACaddressPublisher{}{{Springer, Cham}}.
\newblock
\begin{APACrefURL}
  \url{https://link.springer.com/chapter/10.1007/978-3-319-28028-8_8}
  \end{APACrefURL}
\newblock
\begin{APACrefDOI} \doi{10.1007/978-3-319-28028-8_8} \end{APACrefDOI}
\PrintBackRefs{\CurrentBib}

\bibitem [\protect \citeauthoryear {%
Keller%
}{%
Keller%
}{%
{\protect \APACyear {2007}}%
}]{%
Keller.2007}
\APACinsertmetastar {%
Keller.2007}%
\begin{APACrefauthors}%
Keller, P\BPBI E.%
\end{APACrefauthors}%
\unskip\
\newblock
\APACrefYearMonthDay{2007}{}{}.
\newblock
{\BBOQ}\APACrefatitle {Musical ensemble synchronisation} {Musical ensemble
  synchronisation}.{\BBCQ}
\newblock
\BIn{} \APACrefbtitle {Proceedings of the inaugural international conference on
  music communication science} {Proceedings of the inaugural international
  conference on music communication science}\ (\BPGS\ 80--83).
\PrintBackRefs{\CurrentBib}

\bibitem [\protect \citeauthoryear {%
Kozma%
, Wang%
\BCBL {}\ \BBA {} Zeng%
}{%
Kozma%
\ \protect \BOthers {.}}{%
{\protect \APACyear {2015}}%
}]{%
Kozma.2015}
\APACinsertmetastar {%
Kozma.2015}%
\begin{APACrefauthors}%
Kozma, R.%
, Wang, J.%
\BCBL {}\ \BBA {} Zeng, Z.%
\end{APACrefauthors}%
\unskip\
\newblock
\APACrefYearMonthDay{2015}{}{}.
\newblock
{\BBOQ}\APACrefatitle {Neurodynamics} {Neurodynamics}.{\BBCQ}
\newblock
\BIn{} J.~Kacprzyk\ \BBA {} W.~Pedrycz\ (\BEDS), \APACrefbtitle {Springer
  Handbook of Computational Intelligence} {Springer handbook of computational
  intelligence}\ (\BVOL~58, \BPGS\ 607--648).
\newblock
\APACaddressPublisher{Berlin, Heidelberg}{{Springer Berlin Heidelberg}}.
\PrintBackRefs{\CurrentBib}

\bibitem [\protect \citeauthoryear {%
Linke%
, Bader%
\BCBL {}\ \BBA {} Mores%
}{%
Linke%
\ \protect \BOthers {.}}{%
{\protect \APACyear {2019}}%
{\protect \APACexlab {{\protect \BCnt {1}}}}}]{%
Linke.2019b}
\APACinsertmetastar {%
Linke.2019b}%
\begin{APACrefauthors}%
Linke, S.%
, Bader, R.%
\BCBL {}\ \BBA {} Mores, R.%
\end{APACrefauthors}%
\unskip\
\newblock
\APACrefYearMonthDay{2019{\protect \BCnt {1}}}{}{}.
\newblock
{\BBOQ}\APACrefatitle {The impulse pattern formulation (IPF) as a model of
  musical instruments---Investigation of stability and limits} {The impulse
  pattern formulation (ipf) as a model of musical instruments---investigation
  of stability and limits}.{\BBCQ}
\newblock
\APACjournalVolNumPages{Chaos: An Interdisciplinary Journal of Nonlinear
  Science}{29}{10}{103109}.
\newblock
\begin{APACrefDOI} \doi{10.1063/1.5092511} \end{APACrefDOI}
\PrintBackRefs{\CurrentBib}

\bibitem [\protect \citeauthoryear {%
Linke%
, Bader%
\BCBL {}\ \BBA {} Mores%
}{%
Linke%
\ \protect \BOthers {.}}{%
{\protect \APACyear {2019}}%
{\protect \APACexlab {{\protect \BCnt {2}}}}}]{%
Linke.2019c}
\APACinsertmetastar {%
Linke.2019c}%
\begin{APACrefauthors}%
Linke, S.%
, Bader, R.%
\BCBL {}\ \BBA {} Mores, R.%
\end{APACrefauthors}%
\unskip\
\newblock
\APACrefYearMonthDay{2019{\protect \BCnt {2}}}{}{}.
\newblock
{\BBOQ}\APACrefatitle {The Impulse Pattern Formulation (IPF) as a nonlinear
  model of musical instruments} {The impulse pattern formulation (ipf) as a
  nonlinear model of musical instruments}.{\BBCQ}
\newblock
\BIn{} M.~Kob\ (\BED), \APACrefbtitle {Proceedings of the International
  Symposium on Music Acoustics 2019 - ISMA 2019} {Proceedings of the
  international symposium on music acoustics 2019 - isma 2019}\ (\BPGS\
  336--345).
\newblock
\APACaddressPublisher{Berlin}{}.
\newblock
\begin{APACrefURL}
  \url{http://pub.dega-akustik.de/ISMA2019/data/ISMA_proceedings_all.pdf}
  \end{APACrefURL}
\PrintBackRefs{\CurrentBib}

\bibitem [\protect \citeauthoryear {%
Linke%
, Bader%
\BCBL {}\ \BBA {} Mores%
}{%
Linke%
\ \protect \BOthers {.}}{%
{\protect \APACyear {2021}}%
{\protect \APACexlab {{\protect \BCnt {1}}}}}]{%
Linke.2021}
\APACinsertmetastar {%
Linke.2021}%
\begin{APACrefauthors}%
Linke, S.%
, Bader, R.%
\BCBL {}\ \BBA {} Mores, R.%
\end{APACrefauthors}%
\unskip\
\newblock
\APACrefYearMonthDay{2021{\protect \BCnt {1}}}{}{}.
\newblock
{\BBOQ}\APACrefatitle {Influence of the supporting table on initial transients
  of the fretted zither: An impulse pattern formulation model} {Influence of
  the supporting table on initial transients of the fretted zither: An impulse
  pattern formulation model}.{\BBCQ}
\newblock
\BIn{} \APACrefbtitle {180th Meeting of the Acoustical Society of America}
  {180th meeting of the acoustical society of america}\ (\BPG~035003).
\newblock
\APACaddressPublisher{}{ASA}.
\newblock
\begin{APACrefDOI} \doi{10.1121/2.0001494} \end{APACrefDOI}
\PrintBackRefs{\CurrentBib}

\bibitem [\protect \citeauthoryear {%
Linke%
, Bader%
\BCBL {}\ \BBA {} Mores%
}{%
Linke%
\ \protect \BOthers {.}}{%
{\protect \APACyear {2021}}%
{\protect \APACexlab {{\protect \BCnt {2}}}}}]{%
Linke.2021b}
\APACinsertmetastar {%
Linke.2021b}%
\begin{APACrefauthors}%
Linke, S.%
, Bader, R.%
\BCBL {}\ \BBA {} Mores, R.%
\end{APACrefauthors}%
\unskip\
\newblock
\APACrefYearMonthDay{2021{\protect \BCnt {2}}}{}{}.
\newblock
\APACrefbtitle {Sound examples of an Impulse Pattern Formulation model
  synchronizing to different click tracks.} {Sound examples of an impulse
  pattern formulation model synchronizing to different click tracks.}
\newblock
\APACaddressPublisher{}{Zenodo}.
\newblock
\begin{APACrefDOI} \doi{10.5281/ZENODO.5758991} \end{APACrefDOI}
\PrintBackRefs{\CurrentBib}

\bibitem [\protect \citeauthoryear {%
Michon%
}{%
Michon%
}{%
{\protect \APACyear {1967}}%
}]{%
michon.1967}
\APACinsertmetastar {%
michon.1967}%
\begin{APACrefauthors}%
Michon, J\BPBI A.%
\end{APACrefauthors}%
\unskip\
\newblock
\APACrefYear{1967}.
\newblock
\APACrefbtitle {Timing in temporal tracking} {Timing in temporal tracking}.
\newblock
\APACaddressPublisher{Assen}{{Van Gorcum}}.
\newblock
\begin{APACrefURL}
  \url{http://www.jamichon.nl/jam_writings/1967_ttt_psyfor.pdf}
  \end{APACrefURL}
\PrintBackRefs{\CurrentBib}

\bibitem [\protect \citeauthoryear {%
M{\"u}ller%
, S{\"a}nger%
\BCBL {}\ \BBA {} Lindenberger%
}{%
M{\"u}ller%
\ \protect \BOthers {.}}{%
{\protect \APACyear {2013}}%
}]{%
Muller.2013}
\APACinsertmetastar {%
Muller.2013}%
\begin{APACrefauthors}%
M{\"u}ller, V.%
, S{\"a}nger, J.%
\BCBL {}\ \BBA {} Lindenberger, U.%
\end{APACrefauthors}%
\unskip\
\newblock
\APACrefYearMonthDay{2013}{}{}.
\newblock
{\BBOQ}\APACrefatitle {Intra- and inter-brain synchronization during musical
  improvisation on the guitar} {Intra- and inter-brain synchronization during
  musical improvisation on the guitar}.{\BBCQ}
\newblock
\APACjournalVolNumPages{PloS one}{8}{9}{e73852}.
\newblock
\begin{APACrefDOI} \doi{10.1371/journal.pone.0073852} \end{APACrefDOI}
\PrintBackRefs{\CurrentBib}

\bibitem [\protect \citeauthoryear {%
Parncutt%
}{%
Parncutt%
}{%
{\protect \APACyear {1994}}%
}]{%
Parncutt.1994}
\APACinsertmetastar {%
Parncutt.1994}%
\begin{APACrefauthors}%
Parncutt, R.%
\end{APACrefauthors}%
\unskip\
\newblock
\APACrefYearMonthDay{1994}{}{}.
\newblock
{\BBOQ}\APACrefatitle {A Perceptual Model of Pulse Salience and Metrical Accent
  in Musical Rhythms} {A perceptual model of pulse salience and metrical accent
  in musical rhythms}.{\BBCQ}
\newblock
\APACjournalVolNumPages{Music Perception}{11}{4}{409--464}.
\newblock
\begin{APACrefDOI} \doi{10.2307/40285633} \end{APACrefDOI}
\PrintBackRefs{\CurrentBib}

\bibitem [\protect \citeauthoryear {%
Pikovsky%
, Rosenblum%
\BCBL {}\ \BBA {} Kurths%
}{%
Pikovsky%
\ \protect \BOthers {.}}{%
{\protect \APACyear {2001}}%
}]{%
Pikovsky.2001}
\APACinsertmetastar {%
Pikovsky.2001}%
\begin{APACrefauthors}%
Pikovsky, A.%
, Rosenblum, M.%
\BCBL {}\ \BBA {} Kurths, J.%
\end{APACrefauthors}%
\unskip\
\newblock
\APACrefYear{2001}.
\newblock
\APACrefbtitle {Synchronization: A universal concept in nonlinear sciences}
  {Synchronization: A universal concept in nonlinear sciences}\ (\BVOL~12).
\newblock
\APACaddressPublisher{Cambridge}{{Cambridge University Press}}.
\PrintBackRefs{\CurrentBib}

\bibitem [\protect \citeauthoryear {%
Repp%
}{%
Repp%
}{%
{\protect \APACyear {1992}}%
}]{%
Repp.1992}
\APACinsertmetastar {%
Repp.1992}%
\begin{APACrefauthors}%
Repp, B\BPBI H.%
\end{APACrefauthors}%
\unskip\
\newblock
\APACrefYearMonthDay{1992}{}{}.
\newblock
{\BBOQ}\APACrefatitle {Diversity and commonality in music performance: an
  analysis of timing microstructure in Schumann's Traeumerei} {Diversity and
  commonality in music performance: an analysis of timing microstructure in
  schumann's traeumerei}.{\BBCQ}
\newblock
\APACjournalVolNumPages{The Journal of the Acoustical Society of
  America}{92}{5}{2546--2568}.
\newblock
\begin{APACrefDOI} \doi{10.1121/1.404425} \end{APACrefDOI}
\PrintBackRefs{\CurrentBib}

\bibitem [\protect \citeauthoryear {%
Repp%
}{%
Repp%
}{%
{\protect \APACyear {2001}}%
}]{%
Repp.2001}
\APACinsertmetastar {%
Repp.2001}%
\begin{APACrefauthors}%
Repp, B\BPBI H.%
\end{APACrefauthors}%
\unskip\
\newblock
\APACrefYearMonthDay{2001}{}{}.
\newblock
{\BBOQ}\APACrefatitle {Processes underlying adaptation to tempo changes in
  sensorimotor synchronization} {Processes underlying adaptation to tempo
  changes in sensorimotor synchronization}.{\BBCQ}
\newblock
\APACjournalVolNumPages{Human Movement Science}{20}{3}{277--312}.
\newblock
\begin{APACrefDOI} \doi{10.1016/S0167-9457(01)00049-5} \end{APACrefDOI}
\PrintBackRefs{\CurrentBib}

\bibitem [\protect \citeauthoryear {%
Repp%
}{%
Repp%
}{%
{\protect \APACyear {2005}}%
}]{%
Repp.2005}
\APACinsertmetastar {%
Repp.2005}%
\begin{APACrefauthors}%
Repp, B\BPBI H.%
\end{APACrefauthors}%
\unskip\
\newblock
\APACrefYearMonthDay{2005}{}{}.
\newblock
{\BBOQ}\APACrefatitle {Sensorimotor synchronization: a review of the tapping
  literature} {Sensorimotor synchronization: a review of the tapping
  literature}.{\BBCQ}
\newblock
\APACjournalVolNumPages{Psychonomic Bulletin {\&} Review}{12}{6}{969--992}.
\newblock
\begin{APACrefURL} \url{https://link.springer.com/article/10.3758/bf03206433}
  \end{APACrefURL}
\newblock
\begin{APACrefDOI} \doi{10.3758/BF03206433} \end{APACrefDOI}
\PrintBackRefs{\CurrentBib}

\bibitem [\protect \citeauthoryear {%
Repp%
\ \BBA {} Keller%
}{%
Repp%
\ \BBA {} Keller%
}{%
{\protect \APACyear {2004}}%
}]{%
Repp.2004}
\APACinsertmetastar {%
Repp.2004}%
\begin{APACrefauthors}%
Repp, B\BPBI H.%
\BCBT {}\ \BBA {} Keller, P\BPBI E.%
\end{APACrefauthors}%
\unskip\
\newblock
\APACrefYearMonthDay{2004}{}{}.
\newblock
{\BBOQ}\APACrefatitle {Adaptation to tempo changes in sensorimotor
  synchronization: effects of intention, attention, and awareness} {Adaptation
  to tempo changes in sensorimotor synchronization: effects of intention,
  attention, and awareness}.{\BBCQ}
\newblock
\APACjournalVolNumPages{The Quarterly journal of experimental psychology. A,
  Human experimental psychology}{57}{3}{499--521}.
\newblock
\begin{APACrefDOI} \doi{10.1080/02724980343000369} \end{APACrefDOI}
\PrintBackRefs{\CurrentBib}

\bibitem [\protect \citeauthoryear {%
Shaffer%
}{%
Shaffer%
}{%
{\protect \APACyear {1981}}%
}]{%
Shaffer.1981}
\APACinsertmetastar {%
Shaffer.1981}%
\begin{APACrefauthors}%
Shaffer, L\BPBI H.%
\end{APACrefauthors}%
\unskip\
\newblock
\APACrefYearMonthDay{1981}{}{}.
\newblock
{\BBOQ}\APACrefatitle {Performances of Chopin, Bach, and Bartok: Studies in
  motor programming} {Performances of chopin, bach, and bartok: Studies in
  motor programming}.{\BBCQ}
\newblock
\APACjournalVolNumPages{Cognitive Psychology}{13}{3}{326--376}.
\newblock
\begin{APACrefDOI} \doi{10.1016/0010-0285(81)90013-X} \end{APACrefDOI}
\PrintBackRefs{\CurrentBib}

\bibitem [\protect \citeauthoryear {%
Sogorski%
, Geisel%
\BCBL {}\ \BBA {} Priesemann%
}{%
Sogorski%
\ \protect \BOthers {.}}{%
{\protect \APACyear {2018}}%
}]{%
Sogorski.2018}
\APACinsertmetastar {%
Sogorski.2018}%
\begin{APACrefauthors}%
Sogorski, M.%
, Geisel, T.%
\BCBL {}\ \BBA {} Priesemann, V.%
\end{APACrefauthors}%
\unskip\
\newblock
\APACrefYearMonthDay{2018}{}{}.
\newblock
{\BBOQ}\APACrefatitle {Correlated microtiming deviations in jazz and rock
  music} {Correlated microtiming deviations in jazz and rock music}.{\BBCQ}
\newblock
\APACjournalVolNumPages{PloS one}{13}{1}{e0186361}.
\newblock
\begin{APACrefDOI} \doi{10.1371/journal.pone.0186361} \end{APACrefDOI}
\PrintBackRefs{\CurrentBib}

\bibitem [\protect \citeauthoryear {%
Thaut%
, Miller%
\BCBL {}\ \BBA {} Schauer%
}{%
Thaut%
\ \protect \BOthers {.}}{%
{\protect \APACyear {1998}}%
}]{%
Thaut.1998}
\APACinsertmetastar {%
Thaut.1998}%
\begin{APACrefauthors}%
Thaut, M\BPBI H.%
, Miller, R\BPBI A.%
\BCBL {}\ \BBA {} Schauer, L\BPBI M.%
\end{APACrefauthors}%
\unskip\
\newblock
\APACrefYearMonthDay{1998}{}{}.
\newblock
{\BBOQ}\APACrefatitle {Multiple synchronization strategies in rhythmic
  sensorimotor tasks: phase vs period correction} {Multiple synchronization
  strategies in rhythmic sensorimotor tasks: phase vs period
  correction}.{\BBCQ}
\newblock
\APACjournalVolNumPages{Biological Cybernetics}{79}{3}{241--250}.
\newblock
\begin{APACrefURL}
  \url{https://idp.springer.com/authorize/casa?redirect_uri=https://link.springer.com/article/10.1007/s004220050474&casa_token=be_x34xdjouaaaaa:f2t0tiwtwz3ifj9hl0xj1xidr06qaditz9xeeiuaurxgr70ergo0eh1bfjlogkrtlirk9qrawp3gtjplra}
  \end{APACrefURL}
\newblock
\begin{APACrefDOI} \doi{10.1007/s004220050474} \end{APACrefDOI}
\PrintBackRefs{\CurrentBib}

\bibitem [\protect \citeauthoryear {%
{van Noorden}%
\ \BBA {} Moelants%
}{%
{van Noorden}%
\ \BBA {} Moelants%
}{%
{\protect \APACyear {1999}}%
}]{%
vanNoorden.1999}
\APACinsertmetastar {%
vanNoorden.1999}%
\begin{APACrefauthors}%
{van Noorden}, L.%
\BCBT {}\ \BBA {} Moelants, D.%
\end{APACrefauthors}%
\unskip\
\newblock
\APACrefYearMonthDay{1999}{}{}.
\newblock
{\BBOQ}\APACrefatitle {Resonance in the Perception of Musical Pulse} {Resonance
  in the perception of musical pulse}.{\BBCQ}
\newblock
\APACjournalVolNumPages{Journal of New Music Research}{28}{1}{43--66}.
\newblock
\begin{APACrefDOI} \doi{10.1076/jnmr.28.1.43.3122} \end{APACrefDOI}
\PrintBackRefs{\CurrentBib}

\bibitem [\protect \citeauthoryear {%
{van Orden}%
, Holden%
\BCBL {}\ \BBA {} Turvey%
}{%
{van Orden}%
\ \protect \BOthers {.}}{%
{\protect \APACyear {2003}}%
}]{%
vanOrden.2003}
\APACinsertmetastar {%
vanOrden.2003}%
\begin{APACrefauthors}%
{van Orden}, G\BPBI C.%
, Holden, J\BPBI G.%
\BCBL {}\ \BBA {} Turvey, M\BPBI T.%
\end{APACrefauthors}%
\unskip\
\newblock
\APACrefYearMonthDay{2003}{}{}.
\newblock
{\BBOQ}\APACrefatitle {Self-organization of cognitive performance}
  {Self-organization of cognitive performance}.{\BBCQ}
\newblock
\APACjournalVolNumPages{Journal of experimental psychology.
  General}{132}{3}{331--350}.
\newblock
\begin{APACrefDOI} \doi{10.1037/0096-3445.132.3.331} \end{APACrefDOI}
\PrintBackRefs{\CurrentBib}

\bibitem [\protect \citeauthoryear {%
Wing%
\ \BBA {} Kristofferson%
}{%
Wing%
\ \BBA {} Kristofferson%
}{%
{\protect \APACyear {1973}}%
}]{%
Wing.1973}
\APACinsertmetastar {%
Wing.1973}%
\begin{APACrefauthors}%
Wing, A\BPBI M.%
\BCBT {}\ \BBA {} Kristofferson, A\BPBI B.%
\end{APACrefauthors}%
\unskip\
\newblock
\APACrefYearMonthDay{1973}{}{}.
\newblock
{\BBOQ}\APACrefatitle {Response delays and the timing of discrete motor
  responses} {Response delays and the timing of discrete motor
  responses}.{\BBCQ}
\newblock
\APACjournalVolNumPages{Perception {\&} Psychophysics}{14}{1}{5--12}.
\newblock
\begin{APACrefDOI} \doi{10.3758/BF03198607} \end{APACrefDOI}
\PrintBackRefs{\CurrentBib}

\bibitem [\protect \citeauthoryear {%
Yamada%
}{%
Yamada%
}{%
{\protect \APACyear {1995}}%
}]{%
Yamada.1995}
\APACinsertmetastar {%
Yamada.1995}%
\begin{APACrefauthors}%
Yamada, N.%
\end{APACrefauthors}%
\unskip\
\newblock
\APACrefYearMonthDay{1995}{}{}.
\newblock
{\BBOQ}\APACrefatitle {Nature of variability in rhythmical movement} {Nature of
  variability in rhythmical movement}.{\BBCQ}
\newblock
\APACjournalVolNumPages{Human Movement Science}{14}{3}{371--384}.
\newblock
\begin{APACrefDOI} \doi{10.1016/0167-9457(95)00018-N} \end{APACrefDOI}
\PrintBackRefs{\CurrentBib}

\end{thebibliography}

\end{document}